\documentclass[acmsmall, natbib=false]{acmart}

\AtBeginDocument{%
  }

\copyrightyear{2026}
\acmYear{2026}

\usepackage{subcaption}
\usepackage{algorithm}
\usepackage{algpseudocode}
\def\Comment#1{\textcolor{black}{{#1}}}
\def\Edit#1{#1}

\usepackage{xspace}
\newcommand{\DESIGN}{\mbox{Minos}\xspace}
\newcommand{\insertmultiline}[1]{%
  \begin{tabular}[c]{@{}l@{}}#1\end{tabular}%
}
\acmSubmissionID{44}

\RequirePackage[
  datamodel=acmdatamodel,
  maxnames=200, %
  style=acmnumeric,
  ]{biblatex}

\begin{document}

\title{Minos: Systematically Classifying Performance and Power Characteristics of GPU Workloads on HPC Clusters}

\author{Rutwik Jain}
\authornote{These authors contributed equally to this work and thus are listed alphabetically.}
\affiliation{%
 \institution{University of Wisconsin-Madison}
 \city{Madison}
 \country{USA}}
\email{rnjain@wisc.edu}

\author{Yiwei Jiang}
\authornotemark[1]
\affiliation{%
 \institution{University of Wisconsin-Madison}
 \city{Madison}
 \state{Wisconsin}
 \country{USA}
}
\email{jiang357@wisc.edu}

\author{Matthew D. Sinclair}
\affiliation{%
 \institution{University of Wisconsin-Madison}
 \city{Madison}
 \state{Wisconsin}
 \country{USA}
}
\email{sinclair@cs.wisc.edu}

\author{Shivaram Venkataraman}
\affiliation{%
\institution{University of Wisconsin-Madison}
\city{Madison}
\state{Wisconsin}
\country{USA}}
\email{shivaram@cs.wisc.edu}

\begin{abstract}

As large-scale HPC compute clusters increasingly adopt accelerators such as GPUs to meet the voracious demands of modern workloads, these clusters are increasingly becoming power constrained.
Unfortunately, modern applications can often temporarily exceed the power ratings of the accelerators ("power spikes").
Thus, current and future HPC systems must optimize for both power and performance together.
However, this is made difficult by increasingly diverse applications, which often require bespoke optimizations to run efficiently on each cluster.
Traditionally researchers overcome this problem by profiling applications on specific clusters and optimizing, but the scale, algorithmic diversity, and lack of effective tools make this challenging.
To overcome these inefficiencies, we propose \DESIGN{}, a systematic classification mechanism that identifies similar application characteristics via low-cost profiling for power and performance.
This allows us to group similarly behaving workloads into a finite number of distinct classes and reduce the overhead of extensively profiling 
new workloads.
For example, when predicting frequency capping behavior for a previously unseen application, \DESIGN{} reduces profiling time by 89\%.
Moreover, across 18 popular graph analytics, HPC, HPC+ML, and ML workloads, \DESIGN{} achieves a mean error of 4\% for power predictions and 3\% for performance predictions, significantly improving predictions over state-of-the-art approaches by 10\%.

\end{abstract}


\setcopyright{cc}
\setcctype{by}
\acmJournal{POMACS}
\acmYear{2026} \acmVolume{10} \acmNumber{2} \acmArticle{46}
\acmMonth{6} \acmDOI{10.1145/3805644}

\maketitle

\section{Introduction}
\label{sec:intro}

Modern high performance computing (HPC) clusters increasingly run a diverse set of applications, including graph analytics, HPC, and machine learning (ML).
For example, modern ML algorithms include convolutional neural networks (CNNs), graph neural networks (GNNs), recurrent neural networks (RNNs), Transformers/Large Language Models (LLMs), and Mixtures of Experts (MoEs).
Similarly, HPC applications are also diverse, including %
first principle material calculations~\cite{EISENBACH2017lsms, PhysRevLett-LSMS}, fluid dynamics~\cite{olcf6_mpsdns}, and hydrodynamics~\cite{karlin2013lulesh}.
Some HPC workloads are also being supplemented by ML~\cite{CarterFeddema2023-aiScience, fan2021predicting, jumper2021highly, kates2019predicting} (HPC+ML), including molecular dynamics~\cite{WangZhang2018-deepmd, ZengZhang2023-deepmd2}, %
protein folding~\cite{openfold2}, %
and scientific AI models~\cite{Stevens2023-auroraGPT}. %
As a result, even applications within a specific area (e.g., HPC, ML) can have very different algorithmic properties and system requirements.
Moreover, workload mixes evolve, making it challenging to design and configure HPC systems.

Unfortunately, modern computing systems are also facing tremendous challenges from below due to the slowing of Moore's Law and the end of Dennard's Scaling~\cite{kecklerpicojoule}.
Consequently, to %
meet the computational demands of workloads, HPC systems are increasingly turning towards large-scale clusters of accelerators such as GPUs.
For example, the DOE's Aurora, El Capitan, and Frontier supercomputers have 
37000$-$64000 GPUs.
Further, future HPC systems will likely be comprised of a large variety of compute devices, including new accelerators and customized chips (e.g., TPUs, GraphCore), shared between many users.
Although accelerators are typically more energy efficient than more general-purpose CPUs, they still have large power footprints.
For example, in the past 4 generations, the thermal design power (TDP) of AMD and NVIDIA GPUs has increased from 300 Watts (W)~\cite{mi60, ChoquetteGiroux2018-volta} to 500-750 W~\cite{ampere, LohSchulte2023-mi250} to over \Comment{1400 W}~\cite{AMD_MI355X, Choquette2023-hopper, blackwell, SmithLoh2024-mi300A} per GPU.
Since modern HPC systems often contain tens of thousands of GPUs, and future systems are expected to grow to a million GPUs~\cite{Huang2025-gtcKeynote}, this represents an extremely large -- and growing -- power footprint.
Thus, power management is a major concern for both current and future HPC systems~\cite{FugakuPoints}.

Recently, researchers observed that GPUs used for LLM training/inference in large-scale clusters often temporarily exceed the GPUs rated TDP~\cite{GanRanganathan2025-googlePowerSwings, PatelChoukse2024-POLCA, PigaNarayanan2024-metaDVFS, Sinha-SC22}.
We refer to these variations that exceed the TDP as \textbf{power spikes} because of the instantaneous power spikes in certain GPU kernels (discussed further in Section~\ref{sec:back}).
To combat these power spikes and operate efficiently within constrained power budgets, %
prior work utilizes frequency capping~\cite{PMSysScaleSC24,PatelChoukse2024-POLCA,DLRCap2024} and pinning (discussed further in Section~\ref{sec:back}).
However, these techniques require in-depth, application-specific knowledge of how different workloads are impacted by software-driven frequency settings, and existing work is limited primarily to LLM workloads.
Thus, in modern HPC systems, developers need techniques to help them optimize for performance and power.
Moreover, in Section~\ref{sec:eval-classes} we show that these challenges also apply to diverse workloads from graph analytics, HPC, HPC+ML, and ML.

Traditionally, developers and administrators optimize applications by profiling workloads, then designing algorithmic, hardware, runtime, software, and/or system changes to improve their efficiency on a given system.
\Comment{For example, a system administrator may optimize cluster resource usage by setting frequency caps for jobs running on the cluster.
To do this, they need \textit{apriori} knowledge, typically in the form of power-performance profiles. While such profiles can easily be collected for well known (or benchmark) workloads, if a job that is not part of this pre-profiled set arrives, the sysadmin cannot easily determine optimal frequency capping settings for such jobs.}
However, given the scale, complexity, and number of workloads that are running on modern HPC systems, profiling each workload on each system is increasingly impractical.
Thus, novel research on \emph{workload classification} for HPC systems is required to identify similar workloads such that optimization approaches can be applied across similar workloads without having to carry out expensive profiling per application.

Prior work has recognized the importance of classifying GPU applications~\cite{Aaziz2018proxyapps, Aaziz2019proxyqmcpack, adolf2016fathom, Antici2024MCBound, CoplinBurtscher2016-gpgpuPower, Guerreiro2019dvfs, MITSuperCloud}.
However, these works either observe application performance or power consumption in isolation, without considering their joint impact.
While traditional performance counter-based classification can help drive performance optimizations, we show that the same classification does not hold for power consumption patterns of the same workloads (Section~\ref{sec:eval-classes}).
Thus, performance classification by itself cannot be used for co-designed power and efficiency optimizations, such as power oversubscription using frequency pinning and power capping~\cite{PatelChoukse2024-POLCA} or application power-aware scheduling solutions~\cite{jain2024pal, lettich2024powerfragmentationawareonlinescheduling, Stijkovic25TAPAS}.
We discuss related work further in Section~\ref{sec:related}.

To overcome these issues and enable a more holistic view of GPU workload classification that considers \textbf{both} performance and power, we propose \textbf{\DESIGN{}}.\footnote{\DESIGN{} refers to the chief judge in Greek mythology who sorts humans in the afterlife and is the final arbiter in cases of indecision, much like our scheme classifies workloads based on power- and performance-based clustering.}
Specifically, we propose a novel power classification that uses distributions of a workload's power spikes to group workloads that exhibit similar power patterns and a performance classification that performs clustering based on memory and compute throughput counters (discussed further in Section~\ref{sec:design}; portability and limitations discussed further in Section~\ref{sec:disc}).

Overall, we demonstrate how to \DESIGN{} systematically classify the power and performance behavior for 18 widely-used graph analytics, HPC, HPC+ML, and ML workloads with disparate algorithmic characteristics. 
First, we show how our classification scheme accurately identifies applications from different domains with similar power and performance characteristics.
Then, we use this classification to examine the impact of frequency capping and frequency pinning on workloads from each category: \DESIGN{} accurately predict changes in power and performance scaling behavior for workloads at different frequency caps. 
Next, for previously uncharacterized workloads, we show how \DESIGN{} leverages their performance and power characteristics to accurately predict optimal frequency caps: \DESIGN{} has less than 5\% prediction errors and reduces their profiling overhead by over 90\%.
\DESIGN{} is also generalizable: when applying a hold-one-out approach across 11 unique workloads, \DESIGN{} has a mean absolute prediction error of 4\%,  reducing error by 10\% compared to the state-of-the-art.
To the best of our knowledge, our work is the first to classify GPU applications for both power, performance, and power spikes. 
\Comment{Moreover, we open-source \DESIGN{}'s instructions for running workloads, scripts for profiling at \url{https://github.com/hal-uw/minos-sigmetrics26-artifact}.}

\section{Background}
\label{sec:back} 




Although modern HPC clusters contain thousands or more GPUs, each GPU manages its own power.
Specifically, each GPU has a given thermal limit it must stay within (its TDP).
As mentioned in Section~\ref{sec:intro}, recently AMD and NVIDIA GPUs have increased their TDP by more than 3$\times$ as compute per GPU increased.
The GPU's power management (PM) controller manages the GPU's power consumption. 
As part of this process, the GPU PM controller varies the GPU's voltage and frequency using Dynamic Voltage and Frequency Scaling (DVFS) to avoid exceeding its TDP~\cite{BharadwajDas2024-gpuDVFSPredict, GeVogt2013-dvfsKepler, MeiYung2013-gpuDVFSMeasure, NathTullsen2015-crisp}.
Although GPU vendors 
have not disclosed details about their DVFS schemes (or PM controllers), prior work has shown that, like multi-core CPUs, at some interval (e.g., every $n$ microseconds) DVFS may adjust the GPU's Streaming Multiprocessor (SM) and memory frequencies and voltages to stay within the TDP and reduce energy consumption.
For example, for a GPU kernel that is not very compute intensive, the PM controller will scale the SM frequency and voltage down.
Later, if the PM believes this kernel would benefit from increasing the SM frequency, it will do so.
However, occasionally this process inadvertently compromises performance~\cite{BharadwajDas2024-gpuDVFSPredict, GeVogt2013-dvfsKepler, MeinerzhagenTokunaga2018-gpuDVFS}.

Although the GPU's PM controller 
tries to avoid it, sometimes the workload will temporarily exceed the TDP (a power spike).
The Open Accelerator Infrastructure (OAI) OCP Accelerator Module specification defines the tolerable excursion power of a system as a function of a given socket's TDP~\cite{ocp-spec}.
Although different designs with different requirements are possible, in general the amount by which an accelerator's instantaneous power can exceed the GPU's TDP for a limited amount of time depends on the amplitude of the power excursion.
For instance, a 700W TDP system can draw up to $2\times$ the TDP power limit as long as that excursion does not exceed 20 $\mu$s. 
Thus, different responses are required depending on how long the power spike is allowed to last.
For example, a firmware-based dynamic power manager operating at a frequency of roughly once a millisecond can use DVFS to adjust for longer duration excursions.
However, some excursion time constants (e.g., $2\times$ excursions) are not amenable to firmware management since they must be handled much faster than a millisecond.
Hence, additional hardware-based mechanisms are required to manage higher amplitude excursion violations at a smaller time granularity to guarantee correct operation.
Thus, while temporarily exceeding the TDP is allowed, it makes it difficult to apply prior GPU characterization techniques~\cite{CoplinBurtscher2016-gpgpuPower, Guerreiro2019dvfs} since they cannot easily capture power spikes.
\Edit{While Guerreiro, et al.~\cite{Guerreiro2019dvfs} account for mean power in their methodology, in Section~\ref{subsec:eval-guerreiro} we show they don't account for dynamic power spike variations, allowing \DESIGN{} to outperform it.}

Recent work also explored a number of methods to avoid power spikes.
In the Green 500 methodology~\cite{Scogland2015-pwrPerspectives}, conventional wisdom suggested that system administrators and researchers "pin" the SM's frequency at a specific level to ensure smoother performance -- i.e., to reduce the performance variability of the HPC system~\cite{Sinha-SC22} -- by keeping the frequency constant.
Although the GPU PM can and does overrule this \textbf{frequency pinning} when pinning is used, this typically only happens when the TDP is exceeded.
However, for extremely power constrained systems, as we show in Section~\ref{sec:eval-classes}, this may result in undesirable behavior, including reduced end-to-end performance (Section~\ref{subsec:eval-freq-cap}). 
Accordingly, system designers have started embracing \textbf{frequency capping} to enable large-scale HPC systems to operate more efficiently in extremely power constrained environments~\cite{PMSysScaleSC24,PatelChoukse2024-POLCA,DLRCap2024}.
Unlike frequency pinning, frequency capping does not constrain the SM frequency to a certain value.
Instead, it sets an upper bound on SM frequency and the GPU PM performs DVFS as long as this frequency is not exceeded.
This additional freedom often improves efficiency.
\DESIGN{} is complementary to the specific policy and instead develops a workload classifier, making it easy to deploy policies like frequency capping on diverse applications.



\section{Related Work}
\label{sec:related}



\noindent
\textbf{CPU Workload Classification}:
Prior work has profiled applications using a number of different approaches for CPUs, including SimPoints~\cite{Sherwood01, Sherwood02} and SMARTS~\cite{WunderlichWenisch2003-smarts}.
Although these techniques are still widely used today, they focus on different problems -- how to create a representative subset of a program (e.g., to reduce computer architecture simulation time).
Recently, MCBound classifies CPU HPC jobs as compute- or memory-bound using online profiling augmented by ML~\cite{Antici2024MCBound}.
Like MCBound, we also utilize compute- and memory-intensity.
However, we utilize them for very different purposes -- unlike MCBound, we consider both performance and power, which in Section~\ref{sec:eval-classes} we show is vital to accurately characterize the behavior of modern GPU workloads. 

\noindent
\textbf{GPU Workload Classification}:
Prior work has also classified GPU workloads.
Although these works represent a useful foundation, none of them simultaneously classify applications based on power, power spikes, and performance.
Instead, most classify the similarity between computations~\cite{Aaziz2018proxyapps, Aaziz2019proxyqmcpack, adolf2016fathom}, classify behavior under DVFS~\cite{CoplinBurtscher2016-gpgpuPower, Guerreiro2019dvfs}, or identify subsets of a given workload that are representative of the larger workload~\cite{AvalosKhairy2021-pka, PatiAga20-seqPoints}.
For example, Aaziz et al. studied the relationship between proxy applications and full-scale applications that the proxy applications represent~\cite{Aaziz2018proxyapps, Aaziz2019proxyqmcpack}.
However, we study full-scale applications.
Thus, while Aaziz et al. and \DESIGN{} could be combined, they focus on  different problems.
Prior works~\cite{adolf2016fathom,Aaziz2018proxyapps} utilize performance counters to classify the similarity between workloads and drive optimizations to improve performance.
However, this does not consider 
power consumption.
Other work like AccelWattch~\cite{KandiahPeverelle2021-accelWattch}, GPUWattch~\cite{LengHetherington2013-gpuWattch}, SimplePower~\cite{SimplePower-Vijaykrishnan-ISCA00}, and Wu, et al.~\cite{Gene-GreathouseHPCA15} have developed models to predict GPU performance and/or energy.
However, these models predict performance and power for running applications, not systematically characterizing their combined behavior.
The most relevant prior work to ours is Guerreiro et al. \cite{Guerreiro2019dvfs}, which classifies applications as power- or performance-focused.
However, as we show in Section~\ref{sec:eval-usage}, this classification cannot handle complex power spike behavior in modern applications.

  %
\section{Design}
\label{sec:design}

\DESIGN{} creates a workload classifier that helps uncover workloads with similar behavioral characteristics without requiring expensive system-level profiling for every new application or input per application. 
This classification scheme can then be used to obtain information about a new workload's power consumption, power spikes, and performance to enable intelligent, system-level, application-specific optimizations such as frequency capping or frequency pinning.
To achieve this, \DESIGN{} integrates profiling and clustering to construct a comprehensive view of power and performance scaling for a given workload.
First, we describe how we use low-cost power monitoring data to group workloads based on their power spike distributions.
Second, we leverage GPU kernel granularity hardware metrics and two resource utilization metrics to identify its performance scaling behavior.
Finally, we combine these two classification functions to use \DESIGN{} for frequency capping and pinning optimizations. 

\begin{figure}
    \centering
    \vspace{1ex}
    \includegraphics[width=0.6\linewidth]{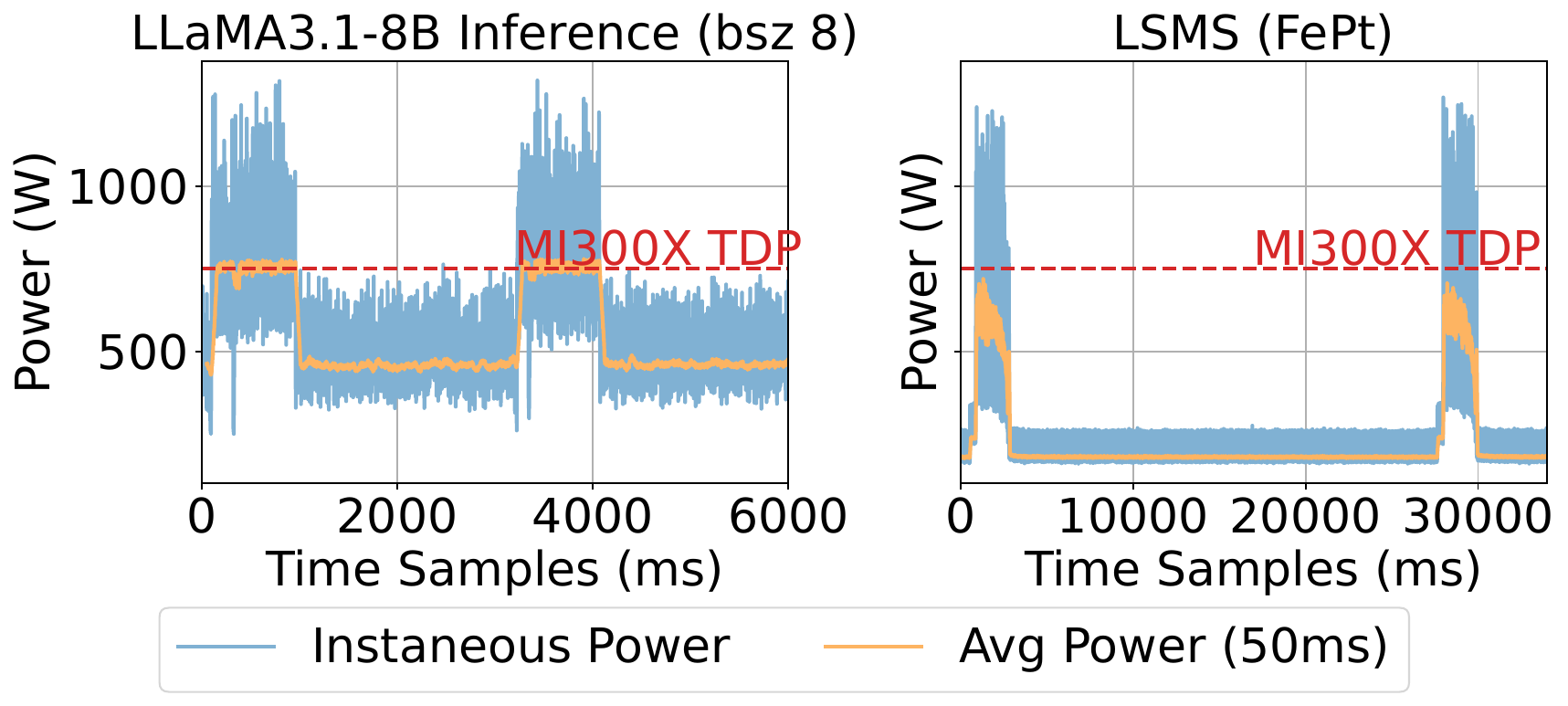}
    \vspace{-2ex}
    \caption{Time series plots showing power behavior for LLaMA3-8B inference and LSMS over two iterations.}
    \Description{Time series plots showing varied power behavior for LLaMA3.1-8B inference and LSMS.}
    \label{fig:pwr-timeseries}
    \vspace{-2ex}
\end{figure}

\subsection{Power-Based Classification}
\label{subsec:design-pwr}

Given that workloads exhibit distinct dynamic power consumption patterns, we first design a classification method that leverages low-overhead power monitoring to capture temporal characteristics valuable for power management and optimization. 
Figure~\ref{fig:pwr-timeseries} shows the power consumption profiles on an AMD MI300X GPU for LLaMA3.1-8B inference (henceforth referred to as LLaMA3) and LSMS, for two iterations of each workload.
First, both workloads have \textbf{power spikes}, which we define as periods where the instantaneous GPU power exceeds the GPU's TDP limit.
These spikes typically occur during transitions from low to high GPU activity, such as the GPU switching from a kernel with low arithmetic intensity to one with high arithmetic intensity.
LLaMA3.1 inference has spikes throughout an iteration run, while LSMS has infrequent bursts of high magnitude spikes with the GPU operating near idle power most of the time ($\approx170$W on the MI300X). 
The magnitude of power spikes and their distribution affects a workload's response to power management optimizations such as frequency capping. 
Systems are also provisioned for peak power consumption, so operators must consider these power spikes when budgeting for system-level power. 

Figure~\ref{fig:pwr-timeseries} shows these workloads's power profiles also exhibit distinct phases. 
LLaMA3.1's compute-intensive prefill phase ($t=0$ to $1000$ms) has very different power behavior than its memory-bound decode phase ($t=1000$ to $3200$ms)~\cite{PatelChoukse2024-POLCA,Patel2024Splitwise}. 
While a LLAMA3.1 iteration might last for a few seconds, LSMS's last tens of seconds.
Thus, LSMS's power fluctuations are less frequent, since only the matrix inversion step is GPU-accelerated, while the rest of the computation 
runs on the CPU, leaving the GPU near idle between bursts.
More broadly, these power variations underscore how a single power metric per workload, whether it be median, average, or peak power consumption, is insufficient to characterize entire applications.

\subsubsection{Feature Extraction}
\label{subsubsec:design-pwr-extract}
To account for these power dynamics over time, as well as the magnitude of power spikes for a given workload, we examine the distribution of power and extract a vector of power usage to better characterize power behavior versus a single power value.
We use the following steps to characterize the distribution and frequency of power spikes by binning them based on magnitude and analyzing their frequency of occurrence:

\begin{enumerate}
    \item \textbf{Spike Detection}: Identify all time samples where instantaneous power $P_{\text{inst}}$ $\geq \text{0.5} \times \text{TDP}$
    
    \item \textbf{Magnitude Computation}: For each identified power spike $i$, compute the relative magnitude regarding TDP: $r_i = P_{\text{inst}}^{(i)}/\text{TDP}$
    
    \item \textbf{Binning}: Define fine-grained bins over the range $[0.5, 2.0]$ with a constant bin width $c$: 
    \begin{equation*}
        [0.5, 0.5+c), [0.5+c, 0.5+2c), \ldots, [2.0-c, 2.0)
    \end{equation*}
    \Edit{We select a lower bound of 0.5 because some applications, such as PageRank, exhibit no power spikes and their vector would be all zeros if we selected 0 as the lower lower. Our upper bound is 2.0 because we do not observe any spikes beyond 2$\times$TDP -- higher-magnitude spikes are suppressed by various PM mechanisms to adhere to the OCP specification~\cite{ocp-spec}.}
    
    \item \textbf{Distribution Vector Construction}: For each bin $[b_j, b_{j+1})$, compute the fraction of power spikes that fall into the bin:
    \begin{equation*}
          v_j = \frac{\# \{ r_i \in [b_j, b_{j+1}) \}}{\text{Total number of spikes}}      
    \end{equation*}
    This results in a normalized vector $\mathbf{v}$ representing the workload’s power spike distribution.
\end{enumerate}

$\mathbf{v}$ encodes how frequently and how significantly each workload exceeds the TDP, thus characterizing its dynamic power profile in the time domain (\Edit{bin size sensitivity evaluated in Section~\ref{subsec:eval-sensitivity}}).

\subsubsection{Clustering}
\label{subsubsec:design-pwr-cluster}

The above feature extraction gives us an $N$ dimensional vector per workload, where $N$ represents the number of power bins.
A smaller $N$ creates coarser grained bins, which are easier to group but may group dissimilar applications together if their coarse, aggregate power values look similar.
Conversely, a larger $N$ creates finer grained bins, enabling \DESIGN{} to distinguish fine-grained power level variations, but also more aggressively separate workloads.
\Comment{To determine the optimal bin granularity, \DESIGN{} uses the \texttt{ChooseBinSize} function in Algorithm~\ref{alg:minos}, a static, low-overhead mechanism that iterates over a small number of bin sizes. We evaluate the sensitivity of bin size on prediction accuracy in Section~\ref{subsec:eval-sensitivity}.}

To identify similarities in power behavior across workloads represented by these vectors, we apply \textbf{Hierarchical Clustering} to the collected feature vectors.
Hierarchical clustering is an agglomerative classification, starting with leaf nodes where every application is its own cluster, and merging similar pairs of clusters together.
To do this, the technique groups clusters using a distance metric. 
Specifically, we use cosine distance to compute pairwise distance between feature vectors since Euclidean distances are biased towards the magnitude of the feature vectors rather than the direction~\cite{cosinedistance,XIA201539}, and cosine similarity does not suffer from this bias.
\Comment{Therefore, we use cosine distance in our evaluation.
Alternative metrics such as Mahalanobis distance~\cite{2018ReprintOM}, which accounts for correlations between features, could potentially capture additional structure in the power spike vectors.}
Like Fathom~\cite{adolf2016fathom}, we also use hierarchical clustering to group workloads with similar dynamic power signatures.

\subsection{Utilization-based Classification}
\label{subsec:design-util}

To mitigate power spikes and lower overall power consumption, power management schemes often cap or pin the GPU's SM frequency (Section~\ref{sec:back}).
However, such schemes must also consider the sensitivity of a workload's performance to the SM frequency and balance power-performance tradeoffs. 
This requires knowledge of the workload's compute and memory sensitivity. 
To enable this, we characterize an application's computation and memory utilization by collecting \textbf{DRAM utilization} ($\text{DRAM}_{\text{Util},ki}$) and \textbf{SM utilization} ($\text{SM}_{\text{Util},ki}$) performance counters for each kernel $i$ ($ki$). 
To derive representative utilization metrics for the entire application, we calculate a weighted average of the per-kernel utilization, using kernel $i$'s \textbf{runtime} ($T_{ki}$) as the weight.
This ensures that longer-running kernels contribute more significantly to the overall application profile.
We compute the application-level DRAM and SM utilization as follows:

\begin{equation}
    \text{App}\,\text{DRAM}_{\text{Util}} = \frac{\sum_{ki} (T_{ki} \times \text{DRAM}_{\text{Util},ki})}{\sum_{ki} (T_{ki})}
\end{equation}

\begin{equation}
    \text{App}\,\text{SM}_{\text{Util}} = \frac{\sum_{ki} (T_{ki} \times \text{SM}_{\text{Util},ki})}{\sum_{ki} (T_{ki})}
\end{equation}
\indent Where $T_{ki}$ is the runtime of kernel $i$.

Next, we represent each application (or input to an application) as a point in the 2-dimensional space defined by its calculated $\text{App}\,\text{DRAM}_{\text{Util}}$ and $\text{App}\,\text{SM}_{\text{Util}}$.
\Comment{For offline analysis and visualization,} we then apply 2-D K-Means clustering on this space to obtain $K_{util}$ number of classes, where $K_{Util}$ denotes the number of utilization-based clusters. 
Based on Silhouette score analysis for $K_{util}$ values ranging from 3 to 17, we determined the optimal $K_{util}$ value (in our case, $K_{util} = 3$\Comment{; details in Section \ref{subsec:eval-classes}}). 
The resulting three clusters group applications into semantically meaningful categories: Compute Intensive, Compute-Memory Hybrid Intensive, and Memory Intensive\Comment{, which helps in understanding utilization patterns across the workload space.
However, \DESIGN{}'s runtime prediction does not rely on clustering results.}
Combining these Power- and Utilization-based classification techniques, we can more holistically characterize a workload. 
Although the power-based classification can help predict a workload's power spikes with different DVFS settings, also using the utilization-based classification allows \DESIGN{} to predict the impact of the same settings on workload performance.

\begin{algorithm}[t]
\footnotesize
\caption{\texttt{SELECT\_OPTIMAL\_FREQ}: Minos Frequency Selection}
\label{alg:minos}
\begin{algorithmic}[1]

\Function{ChooseBinSize}{$T$, $P_T$, $E_f$, $\mathcal{C}$}
    \State \Return $\displaystyle \arg\min_{c \in \mathcal{C}} \textsc{P90PwrPredErr}(T, \textsc{GetPwrNeighbor}(T, P_T, E_f, c))$
\EndFunction 

\Function{GetPwrNeighbor}{$T$, $P_T$, $E_f$, \Edit{$c$}}
    \State Construct power spike vector $\mathbf{v}$ \Edit{for bin size $c$}
    \State Add $T$ to dendrogram constructed from $E_f$ using $\mathbf{v}$
    \State Find reference app $R_{pwr} \gets \arg\min_R \text{cosine\_distance}(\mathbf{v}, \mathbf{v}_R)$
    \State \Return $R_{pwr}$
\EndFunction

\Function{GetUtilNeighbor}{$T$, $App\,SM_{Util}, App\,DRAM_{Util}, E_f$}
    \State Add $T$ to K-Means clustering of utilization data from $E_f$
    \State Identify reference app $R_{util}$ from the same cluster as $T$
    \State \Return $R_{util}$
\EndFunction

\Function{CapPowerCentric}{$R_{pwr}$}
    \State $\textbf{bound} \gets 1.3 \times \text{TDP}$
    \ForAll{frequency $f$ in decreasing order from $R_{pwr}$'s scaling data}
        \If{$P^{90\%}_{R_{pwr}, f} < \textit{bound}$}
            \State \Return $f$
        \EndIf
    \EndFor
\EndFunction

\Function{CapPerfCentric}{$R_{util}$}
    \State $\textbf{bound} \gets 5\%$ degradation
    \ForAll{frequency $f$ in increasing order from $R_{util}$'s scaling data}
        \If{perf. degradation $\leq \textit{bound}$}
            \State \Return $f$
        \EndIf
    \EndFor
\EndFunction

\Function{Main}{}
    \State $c^{*} \gets$ \Call{ChooseBinSize}{$T, P_T, E_f, \mathcal{C}$}
    \State $R_{pwr} \gets$ \Call{GetPwrNeighbor}{$T, P_T, E_f, c^{*}$}
    \State $R_{util} \gets$ \Call{GetUtilNeighbor}{$T, App\,SM_{Util}, App\,DRAM_{Util}, E_f$}
    \State $f_{pwr} \gets$ \Call{CapPowerCentric}{$R_{pwr}$}
    \State $f_{perf} \gets$ \Call{CapPerfCentric}{$R_{util}$}
    \State $f_{cap} \gets f_{pwr}$ or $f_{perf}$ depending on objective
    \State \Return $f_{cap}$
\EndFunction

\end{algorithmic}
\end{algorithm}

\subsection{Using \DESIGN{}}
\label{subsec:design-usingminos}

Our classification can inform (a) sophisticated power management strategies, such as workload-specific DVFS settings, frequency or power capping, and (b) enable more efficient resource allocation and placement decisions within cluster scheduling frameworks.
\Comment{System administrators are increasingly moving towards profile-guided optimizations to limit jobs' energy consumption, improve overall resource utilization, and increase system reliability.
All of these are directly affected by workload power behavior, particularly power spikes. 
Several job schedulers, such as POLCA~\cite{PatelChoukse2024-POLCA}, TAPAS~\cite{Stijkovic25TAPAS}, and PAL~\cite{jain2024pal}, perform workload-specific optimizations to achieve better resource utilization. 
Properly characterizing how workloads on systems with varying power and frequency limits is essential to utilize these optimizations.}
Moreover, modern HPC clusters run a large, diverse set of graph analytics, HPC, HPC+ML, and ML workloads, with different behaviors and characteristics~\cite{openfold2, DingZheng2023-mirage, olcf6_benchmarks, mlcommons_benchmarks, nersc10, ZengZhang2023-deepmd2}.
For example, for workloads with strict latency or SLO (Service Level Objective) restrictions, such as conversational LLM inference, 
we must strictly bound the performance loss due to frequency capping so that it does not violate SLOs. 
We call this \textsc{PerfCentric} optimal frequency selection. 
Conversely, for workloads with more relaxed SLOs, such as ML training, LLM summarization tasks, and scientific simulations such as LAMMPS, we can select a strict bound on the power spikes of the workload, potentially tolerating a performance degradation.
We call frequency selection for such scenarios \textsc{PowerCentric}. 
Moreover, these workloads often have multiple input datasets.
In Section~\ref{sec:eval-classes} we show that these datasets sometimes result in very different performance and power characteristics for a given application.
Consequently, profiling each workload, and each input for that workload, to identify the appropriate application- or input-specific strategies for power management optimizations is time consuming and challenging. 
In Section~\ref{subsec:eval-case-study} we show how \DESIGN{} identifies optimal frequency caps for new workloads using low-cost profiling and \DESIGN{}'s classification.

Since frequency capping manages power-performance tradeoffs, the optimal capping decision is different for a \textsc{PerfCentric} approach versus a \textsc{PowerCentric} approach. 
Algorithm~\ref{alg:minos} details the steps to select the optimal frequency cap under either condition using \DESIGN{}'s classification. \Comment{\textsc{ChooseBinSize} is a lightweight, one-time, offline step that searches over a small candidate set to select the bin size that minimizes prediction error.
Since \textsc{ChooseBinSize} occurs offline, it does not incur any runtime overhead.}
The key idea is that, given a new target workload, we add this workload to the power- and utilization-based classifications that \DESIGN{} produces and identify the nearest neighbor to this workload. 
\textsc{GetPwrNeighbor} and \textsc{GetUtilNeighbor} functions use the power-based and utilization-based classifications in Sections~\ref{subsec:design-pwr} and~\ref{subsec:design-util} respectively to identify the nearest neighbor applications to the target workload.
The performance and frequency scaling data of the nearest neighbors can then be used to predict how the new workload would react under different frequency caps. 
For the \textsc{PowerCentric} approach (\textsc{CapPowerCentric}), we  want to find an optimal frequency cap for a workload which ensures that the 90th percentile power spikes do not exceed a bound (we use $1.3\times$TDP as the bound in our evaluation). 
To do so, we use the power neighbor's frequency scaling data to find the highest possible frequency cap at which the neighbor's p90 power spikes are within the specified bound. 
This provides the predicted optimal frequency $f_{pwr}$ for the target workload. 
For the \textsc{PerfCentric} approach (\textsc{CapPerfCentric}), we want to implement a strict performance bound (5\% performance degradation in our evaluation). 
Our algorithm observes the frequency scaling behavior of the performance neighbor $R_{perf}$ to find the lowest possible frequency where the performance degradation is limited to within the specified bound. 
This frequency is selected as the optimal frequency $f_{perf}$.
We show how our dual-classification scheme can be jointly used for optimal frequency capping in Sections~\ref{subsec:eval-case-study} and~\ref{subsec:eval-pairwise}.

{ \small
  \centering
\begin{table*}[th!]
  \caption{Workloads used in our classification (* indicates workloads run with reduced size due to cluster runtime limits).}
  \label{tab:workloads-util}
    \begin{tabular}{llllll}
    \toprule
    
    \textbf{Workload}  &
    \textbf{Domain}    & 
    \insertmultiline{\textbf{Model/}\\\textbf{Impl}} 
    & \textbf{Config/Inputs} 
    & \textbf{PwrClass} 
    & \textbf{PerfClass} \\ 

    \midrule
    
    SGEMM~\cite{cublas}     &  
    $\mu$benchmark          & 
    cublasSgemm             &
    $25536 \times 25536$    &
    -                       &   
    C5                      \\
    \midrule
    
    \insertmultiline{PageRank\\\cite{CheBeckmann2013-pannotia,PageRankSpMV,Wang2017Gunrock}}  & 
    \insertmultiline{Graph \\ Analytics}                                    & 
    \insertmultiline{Pannotia,\\Gunrock}          &    
    \insertmultiline{indochina\\ at\&t  \\ }               &   
    \insertmultiline{Low-spike \\ Low-spike}                    &
    \insertmultiline{H6,M3\\ C1,C4}                 \\
    
    \insertmultiline{
        BFS~\cite{Wang2017Gunrock}\\ 
        SSSP~\cite{Wang2017Gunrock}\\
        BC~\cite{Wang2017Gunrock}
    }                                                                       &
    \insertmultiline{Graph \\ Analytics}                                    & 
    Gunrock                                                                 & 
    \insertmultiline{indochina\\ kron}                                      &
    -                                                                       &  
    \insertmultiline{M5,M8,M7\\ M4,M10,M6} \\
    \midrule
    
    LULESH~\cite{karlin2013lulesh}                                          & 
    HPC                                                                     &  
    v2.0                                                                    & 
    \insertmultiline{n 300 i 10 \\ n 500 i 10}      &    
    \insertmultiline{Mixed \\ High-spike}        &  
    H5 \\
    
    LSMS~\cite{PhysRevLett-LSMS,EISENBACH2017lsms}                          &  
    HPC                                                                     & 
   -                                                                     & 
    FePt,lmax=5,rLIZ=18                                                   &  
    Mixed &  
    M1*\\
    
    LAMMPS~\cite{LAMMPS}                                                    &
    HPC                                                                     &
    in.eam                                                                  & 
    \insertmultiline{$(8,8,16)$ \\  $(16,16,16)$}           & 
    \insertmultiline{High-spike \\High-spike}                               &
    C3         \\
    
    
    MILC~\cite{olcf6_milc}                                                  &
    HPC                                                                     &
    \insertmultiline{su3\_rhmd\\\_hisq}                                                         &  \insertmultiline{$24\times24\times24\times6$ \\$6\times6\times6\times6$ }                                                &  
    Low-spike  &
    H4,M2               \\
    
    M-PSDNS~\cite{olcf6_mpsdns}                                             &
    HPC                                                                     &
    -                                                                       & $990\times990\times990$ FP32                                            & 
    -                                                                       &
    C8 \\
    
    \midrule
    
    \insertmultiline{LLaMA2 \\ Training~\cite{touvron2023llama2openfoundation}}                  &
    ML                                                                      &
    \insertmultiline{LLaMA2-7B\\torchtune}                        &
    \insertmultiline{dataset: alpaca \\bsz 32, 64}                 &            
    Mixed                                                                  &
    M9*    \\
    
    \insertmultiline{LLaMA2 \\ Inference~\cite{touvron2023llama2openfoundation}}               & 
    ML                                                                      & 
    \insertmultiline{LLaMA2-7B\\vLLM}                                      &
    bsz 1,8,32                                    & 
    \insertmultiline{Mixed \\ High-spike}                                  & 
    C7* \\
    
    \insertmultiline{LLaMA3\\Inference~\cite{grattafiori2024llama3herdmodels}}                 &
    ML                                                                      &
    \insertmultiline{LLaMA3.1-8B\\vLLM}                                   & 
        bsz 1,8,32                                 & 
    \insertmultiline{Low-spike \\ High-spike}                               &
    H1* \\
    
    \insertmultiline{Stable \\Diffusion(SD-XL)~\cite{podell2023sdxlimprovinglatentdiffusion}}         &
    ML                                                                      &
    SDXL Turbo                                                              &    
    \insertmultiline{bsz 16,32\\res:512,1K,2K}                                                                &
    High-spike                                                              &
    -                   \\     
    
    GNN~\cite{chen2021rgatrelationalgraphattention}                         &
    ML                                                                      & 
    r-GAT &   
    \insertmultiline{IGBH-tiny \\ bsz 1024,steps 5 }                &
    -  &
    C6 \\
    
    ResNet50~\cite{ResNet-pyTorchRef}                                       &
    ML                                                                      & 
    torchvision                                                             &
          \insertmultiline{ImageNet \\ CIFAR-10} \insertmultiline{bsz 256,512}          
          &  Mixed
                                                   &
    H2 \\
    
    \midrule
   
    DeePMD~\cite{ZengZhang2023-deepmd2}                                     & 
    HPC + ML                                                                &   
                \insertmultiline{Water\\ DPA2} & 
    \insertmultiline{bsz 32,64,128,auto}             & 
    Mixed                                                                 &
     \insertmultiline{C9 *\\ H3 *}      \\
    
    OpenFold~\cite{Ahdritz2022openfold,ahdritz2023openproteinset}           &
    HPC + ML                                                                &
    \insertmultiline{MLCommons\\ AQLaboratory}                              &
    \insertmultiline{OpenProteinSet \\ bsz 1,2,4,8}           &
    Mixed & C2  \\    
    \bottomrule
    \end{tabular}

\end{table*}

}

\section{Methodology}
\label{sec:meth}

\subsection{Clusters}
\label{subsec:meth-clusters}

To validate the utility and effectiveness of our classification scheme, we ran experiments on two high-performance computing clusters with different GPU vendors. 
Specifically, we use AMD's \textbf{HPC Fund} cluster and the \textbf{Lonestar6 cluster} at the Texas Advanced Computing Center (TACC).
The HPC Fund cluster contains MI300X nodes, with 8 MI300X GPUs per node and 192 GB of HBM memory. 
The LoneStar6 cluster has NVIDIA GPUs, with each node having 3 A100 PCIe GPUs with 40 GB of HBM2 memory.

For the power-based classification, we utilize AMD's 
ROCm System Management Interface (\texttt{rsmi}) API~\cite{rsmiapi}, to collect power data at a 1-2 millisecond granularity. 
Collecting similar data for the NVIDIA GPUs required administrative privileges for frequency capping and pinning, which we did not have on Lonestar6. 
We did perform utilization-based classification 
on NVIDIA GPUs (using SM throughput and DRAM throughput) in LoneStar6. 
However, since the AMD and NVIDIA profilers do not measure utilization the same way, we do not directly compare them.
\Comment{Although our evaluation uses AMD and NVIDIA GPUs, \DESIGN{} only requires power telemetry and utilization counters, which are available on all modern GPUs.}

\subsection{Workloads}
\label{subsec:meth-apps}

To ensure that our classification captures a wide range of application behavior, we profile a diverse set of benchmarks that stress different GPU components.
Table~\ref{tab:workloads-util} summarizes our workloads, configurations, and inputs.
Table~\ref{tab:workloads-util} also indicates the power- and utilization-based classes that each workload gets categorized into (discussed further in Sections~\ref{subsubsec:meth-util-class} and \ref{sec:eval-classes}).
These workloads include applications from graph analytics, HPC, HPC+ML, and ML workloads from popular benchmark suites like CORAL-2~\cite{llnl_coral2_benchmarks}, Gunrock~\cite{Wang2017Gunrock}, MLPerf~\cite{mlcommons_benchmarks}, OLCF-6 ~\cite{olcf6_benchmarks}, and Pannotia~\cite{CheBeckmann2013-pannotia}. 

Our HPC workloads (e.g., from CORAL-2 and OLCF-6) span various scientific computing domains, including first principle calculation, hydrodynamics, molecular dynamics, and quantum chromodynamics.
We selected these workloads because they stress various GPU components, including memory bandwidth, floating point performance, and memory latency. 
Graph analysis is also widely utilized in modern HPC systems.
We ran PageRank, a graph analytics benchmark with different graphs, and also used two distinct implementations from Pannotia~\cite{CheBeckmann2013-pannotia} and Gunrock~\cite{Wang2017Gunrock}. 
Additionally, we also evaluated Gunrock's Betweenness Centrality (BC), Single-Source Shortest Path (SSSP), and Breadth-First Search (BFS) workloads with various input graphs of different sizes.
Collectively, this allows us to observe input- and implementation-dependent variation in characteristics across these important workloads.

We also classify several ML workloads: ResNet50 training (a popular image classification model), R-GAT (a graph attention network), LLaMA2-7B Large Language Model (LLM) training, and LLaMA2-7B and LLaMA3.1-8B inference.
Finally, we also examine hybrid models that combine ML with scientific computing; we examine two workloads that introduce ML into traditional HPC simulation. 
DeePMD~\cite{ZengZhang2023-deepmd2} uses deep learning to help solve molecular dynamics problems, while OpenFold~\cite{Ahdritz2022openfold} uses an inference pipeline to predict protein folding.
We selected application batch and input sizes to stress different GPU compute and memory utilizations while also ensuring they fit on the GPU (Table~\ref{tab:workloads-util}).

\subsection{Profiling and Post-Processing}
\label{subsec:meth-profiling}

\begin{figure}[tb!]
    \centering
    \includegraphics[width=0.7\linewidth]{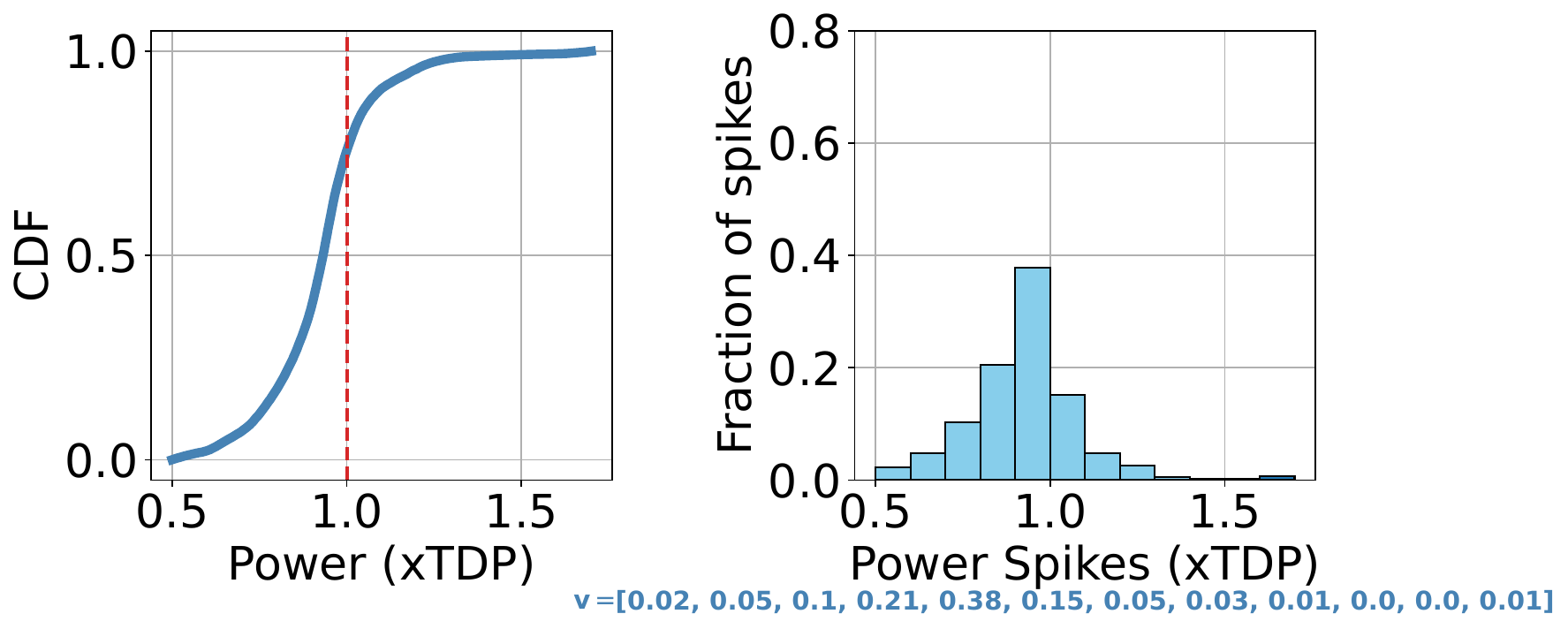}
    \vspace{-2ex}
    \caption{Cumulative power spike distribution for LLaMA3-8B inference (left) and histogram showing fraction of spikes if binning is performed with bin size = 0.1 (right), along with the resultant power spike vector \textbf{v}.}
    \Description{Cumulative distribution of power spikes for LLaMA3-8B inference (left) and histogram showing fraction of spikes if binning is performed with bin size =  0.1 (right) }
    \label{fig:power-methodology}
    \vspace{-4ex}
\end{figure}

\subsubsection{Profiling Power Consumption}
\label{subsubsec:meth-profiling-power}

\Comment{\DESIGN{}'s profiling only relies on vendor-provided power monitoring interfaces, such as AMD's \texttt{rsmi} or NVIDIA's \texttt{NVML}.}
On AMD GPUs we use \texttt{rsmi} to collect application telemetry data.
Specifically, we periodically queried hardware counters using the following API calls:

\begin{itemize}
    \item \texttt{rsmi\_dev\_power\_ave\_get()}: Returns power consumption of the GPU device in $\mu W$.
    \item \texttt{rsmi\_dev\_energy\_count\_get()}: Records the energy accumulator counter of the GPU device between the last sample and the current sample. 
\end{itemize}

We designed a low-overhead wrapper over the RSMI API that provides $\approx$1-2 ms granularity samples. 
However, we found the RSMI API provides a heavily averaged power value when using \texttt{power\_ave\_get()}, which is filtered or averaged over multiple milliseconds. 
Thus, to get a more faithful instantaneous power consumption $P_{inst}$ measurement, we use the change in the accumulated energy counter ($\Delta e$) between successive samples ($\Delta t$): $P_{inst} \approx {\Delta e}/{\Delta t}$. 

Conversely, power derived from AMD's energy counter was too noisy, with high spikes~\cite{YangNvidiaPowerSensor}.
To mitigate high-frequency noise in this derived instantaneous power, we applied an exponential moving average filter (alpha filter) with a coefficient $\alpha = 0.5$.
We chose $0.5$ because it smooths out noisy outlier values by performing successive-sample averaging:
\begin{align*}
    P_{filt}(t) &= \alpha P_{inst}(t) + (1 - \alpha)P_{inst}(t -1) \\
                &=  (P_{inst}(t) + P_{inst}(t -1))/2 , \,\, \alpha = 0.5
\end{align*}

To accurately capture the relevant power trace corresponding to active GPU execution, we monitored the \texttt{SQ\_BUSY\_CYCLES} counter (indicating when AMD CUs are active).
We filter the trace to only contain records from when this counter first and last indicated non-zero activity, trimming idle periods at the beginning and end of the trace.

\subsubsection{Classification Method}
\label{subsubsec:meth-power-class}

From the filtered power profiles, we extracted features to characterize each workload’s dynamic power behavior.
As outlined in Section~\ref{subsec:design-pwr}, we constructed a per-application vector that represents the distribution of its power relative to TDP.
Figure~\ref{fig:power-methodology} demonstrates this for LLaMA3.1-8B inference, with a $0.1\times$ TDP bin size.
By default our results also use a bin size of $c = 0.1\times$ TDP to capture broad CDF patterns when grouping workloads, while still providing sufficiently fine-grained power visibility.
\Comment{In Section~\ref{subsec:eval-sensitivity} we evaluate bin size sensitivity.}

We then applied hierarchical clustering on these spike distribution vectors, using ward linkage and cosine distance~\cite{mullner2011modernhierarchicalagglomerativeclustering}.
The resulting dendrogram groups workloads with similar power spike distributions. 
We also sliced the dendrogram for purposes of explanation in Section~\ref{sec:eval-classes} to obtain $K_{power}$ classes, where $K_{power} >= 2$ and $K_{power}$ is less than the number of applications.
In Section~\ref{sec:eval-classes} we analyze $K_{power} = 3$ classes.
\Comment{Note that the dendogram and slicing threshold does not affect \DESIGN{}'s predictions, since Algorithm~\ref{alg:minos} only uses the nearest neighbor rather than dendogram classes.
The slicing threshold is only used for understanding \DESIGN{}'s classification in Section~\ref{sec:eval-classes}, not in \DESIGN{}'s actual deployment in Section~\ref{sec:eval-usage}.}

\subsubsection{Frequency Capping}
\label{subsec:meth-freq-capping}

To demonstrate \DESIGN{}'s validity and utility, we perform a case study of how we can introduce frequency capping for new workloads. 
However, since we only had administrative privileges on the MI300X GPUs, we ran our frequency capping experiments on these GPUs. 
We sweep values for the SM or CU frequency cap from 1300 MHz to 2100 MHz (the boost frequency on MI300X devices), where 2100 MHz represents the uncapped behavior.
At each of these frequencies, we profile the power consumption distribution using the RSMI API, as well as performance metrics, such as execution time or iteration time recorded by the application. 
Sections~\ref{subsec:eval-freq-cap} and~\ref{subsec:eval-case-study} present results from our case study.

\subsubsection{Profiling Utilization}
\label{subsubsec:meth-util-prof}

We employed NVIDIA's \texttt{nsight} \texttt{compute}~\cite{nvidia_nsight_compute} to gather key hardware performance metrics to characterize the resource utilization of applications in Table~\ref{tab:workloads-util}.
Specifically, for all GPU kernels executed within each application run, we collected the following performance counters:

\begin{enumerate}
    \item \textbf{DRAM Throughput}: Measured using 
    \texttt{gpu\_\_dram\_throughput.avg.pct\_of\_peak\_\\sustained\_elapsed}, the percent utilization of peak sustained memory bandwidth.
    \item \textbf{SM Throughput}: Measured using 
    \texttt{sm\_\_throughput.avg.pct\_of\_peak\_sustained\_\\elapsed}, representing utilization percentage of peak sustained SM compute throughput.
  
    \item \textbf{Kernel Duration}: Obtained using the 
    \texttt{gpu\_\_time\_duration.sum} metric (labeled as "Duration"), providing the execution time ($T_{ki}$) for each kernel $ki$.
\end{enumerate}

To minimize GPU profiling overheads and ensure that the collected data accurately reflects the core computational behavior on the GPU, we profile only the application's main code loop(s).
To achieve this, we inserted \texttt{profiler.start()} and \texttt{profiler.end()} API calls around these 
source code regions. 
This allowed us to precisely capture the performance characteristics of representative application parts while significantly reducing the time and data volume associated with profiling.

While we focus on utilization-based classification using NVIDIA GPUs, AMD GPUs also expose similar performance metrics -- such as CU activity and memory bandwidth utilization -- via ROCm's profiling tools (e.g., \texttt{rocprof}). 

\subsubsection{Classification Method for Utilization}
\label{subsubsec:meth-util-class}

As discussed in Section~\ref{sec:design}, we calculated a kernel duration weighted average of the per-kernel metrics.
This ensures longer-running kernels contribute proportionally more to the overall utilization. 
Consequently, we represent each application as a point in the two-dimensional space defined by its calculated $\text{App}\,\text{DRAM}_{Util}$ and $\text{App}\,\text{SM}_{Util}$.
\Comment{Similar to the power-based classification, this 2-D utilization representation serves two purposes.
At runtime, Minos uses it to find the nearest neighbor and predict performance sensitivity using the closest neighboring workload; offline we apply K-Means clustering (with Silhouette-based $K$ selection) to derive coarse utilization categories for interpretability.}

\section{Understanding \DESIGN's Classification}
\label{sec:eval-classes}

In this section, we show that \DESIGN{} effectively groups similarly behaving workloads into distinct classes (Section~\ref{subsec:eval-classes}), including when applying frequency capping or pinning (Section~\ref{subsec:eval-freq-cap}).
Interestingly, different inputs for a given workload can fall into different classes when they significantly differ in compute and memory utilization.
Moreover, although capping or pinning the frequency affects the performance and power spikes, \DESIGN{} effectively captures scaling trends in its grouping (Section~\ref{subsec:eval-freq-cap}).
Section~\ref{sec:eval-usage} evaluates how \DESIGN{} can be used for unseen workloads to determine optimal frequency cap settings, showing that \DESIGN{} can accurately predict 90th percentile power spikes (4\% average error across workloads) and performance (3\% average error). 


\subsection{Classification Results}
\label{subsec:eval-classes}

\begin{figure}
    \centering
    \vspace{1ex}
    \includegraphics[width=0.9\linewidth]{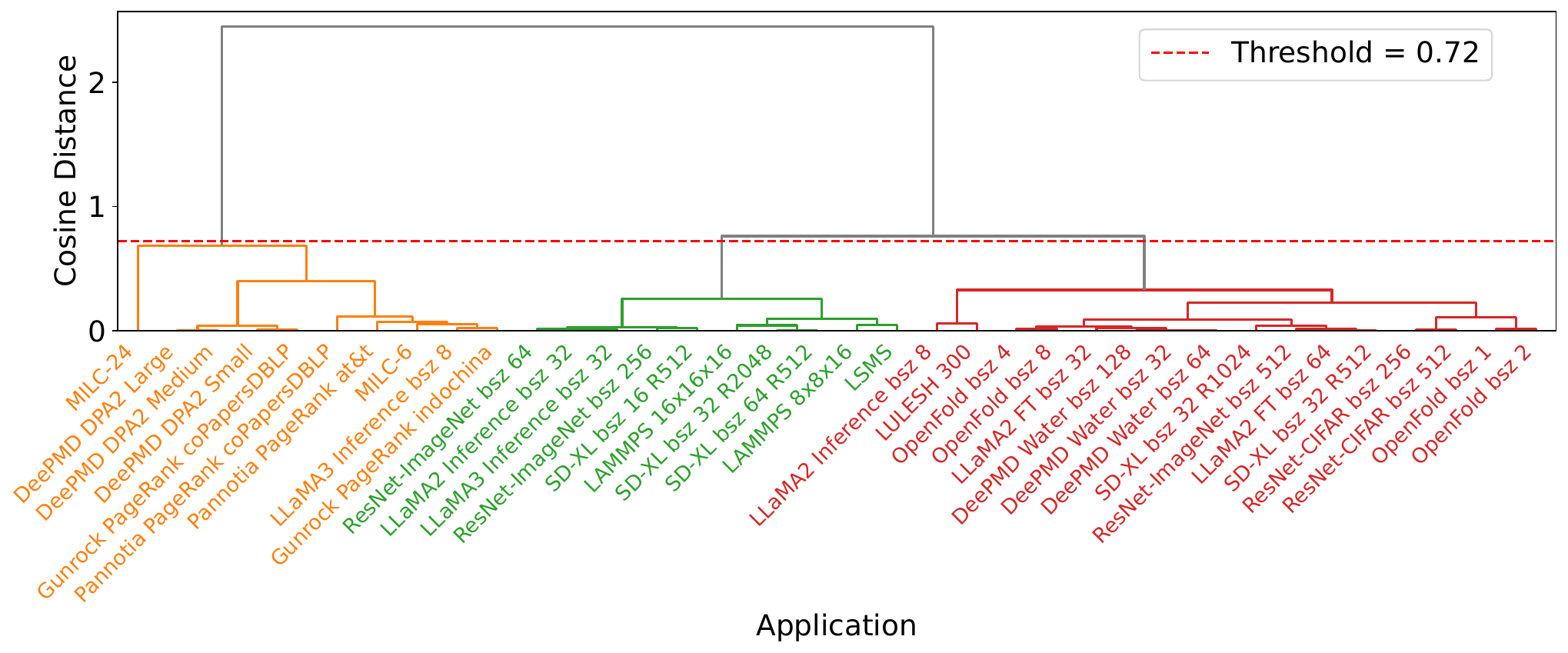}
    \vspace{-3ex}
    \caption{Dendrogram based on power spike distributions of workloads. We label the clusters as Low-spike (orange), High-spike (green), and Mixed (red), respectively, based on their power distribution.}
    \Description{Dendrogram showing hierarchical clustering based on power spike distributions of workloads.}
    \label{fig:power-cluster}
    \vspace{-1ex}
\end{figure}

We use power spike distributions as the primary categorization method and incorporate resource utilization patterns to understand the underlying causes of different power behaviors.
Figure~\ref{fig:power-cluster} shows the dendrogram 
based on their power distribution vectors. 
The dendrogram's y-axis indicates the cosine distance between two workloads.
A cosine distance of 0 indicates perfectly aligned vectors, while a larger cosine distance indicates that workloads are farther apart.
Thus, we can slice the dendrogram at suitable cosine distances to obtain $K$ different groups or clusters.
Slicing the dendrogram at a cosine distance of 0.72 yields three distinct power behavior groups ($K$=3).
Figure~\ref{fig:cdf_comparison} illustrates the cumulative power distributions for these groups.
The exact choice of the threshold (and number of groups of $K$) does not affect \Comment{\DESIGN{}'s predictions}, because \DESIGN{}'s Algorithm~\ref{alg:minos} uses each application's nearest neighbor as the predictor rather than cluster labels.
We show these three clusters for simplicity, to better group the applications, and to explain their similar behavior trend within each class.

Based on the distinct distribution shapes for the three groups, we label them as Low-spike (orange in Figure~\ref{fig:power-cluster}), High-spike (green) and Mixed workloads (red).

\begin{figure}
    \centering
    \includegraphics[width=1\linewidth]{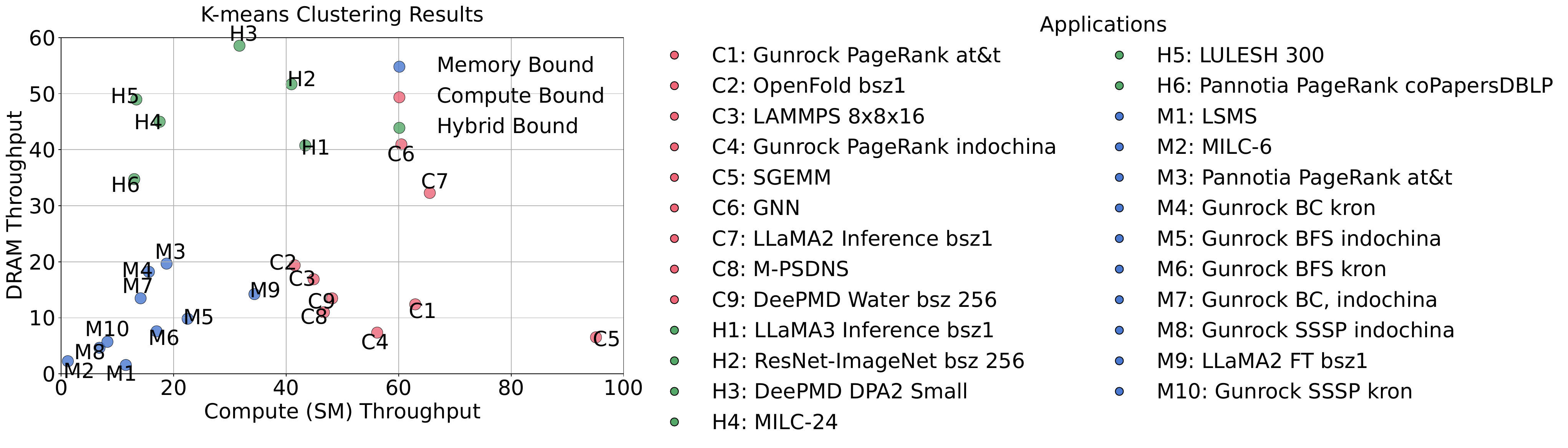}
    \vspace{-4ex}
    \caption{K-Means Clustering on memory and compute utilization showing workloads grouped as C (compute-intensive), M (memory-intensive) and H (hybrid).}
    \Description{Clustering on memory and compute utilization showing workloads grouped into $K_{util}=3$ classes,  C (Compute-intensive), M (memory-intensive) and H (hybrid).}
    \label{fig:util-cluster}
    \vspace{-2ex}
\end{figure}

In addition to this dendrogram, Figure~\ref{fig:util-cluster} shows how the workloads cluster in terms of resource utilization.
For the K-Means clustering, we applied a silhouette score analysis by sweeping $K_{util}$ values from 3 to 17.
Since \textbf{$K_{util}$=3} yielded the highest Silhouette Score ($0.48$), we use three clusters.
Based on their utilization values, we logically label these clusters as compute-intensive (C class), memory-intensive (M class) and Hybrid workloads (H class). 
Compute-intensive workloads tend to exhibit low DRAM throughput (usually below 15\%) while showing a broad range of SM throughput ($\approx$40–95\%).
In contrast, Memory-bound applications display low SM throughput (generally below 40\%, often under 20\%) but show greater variability in DRAM throughput (ranging from $\approx$10\% to 55\%). 
Hybrid applications occupy an intermediate region with a more balanced utilization of both compute and memory resources. 
We highlight three key insights from this data.

\subsubsection{Power Spike Characteristics Influenced by Resource Utilization}
\label{subsubsec:pwr-util-correlation}

\begin{figure}[tb!]
    \centering
    \vspace{1ex}
    \begin{subfigure}[b]{0.33\textwidth}
        \centering
        \includegraphics[width=0.9\textwidth]{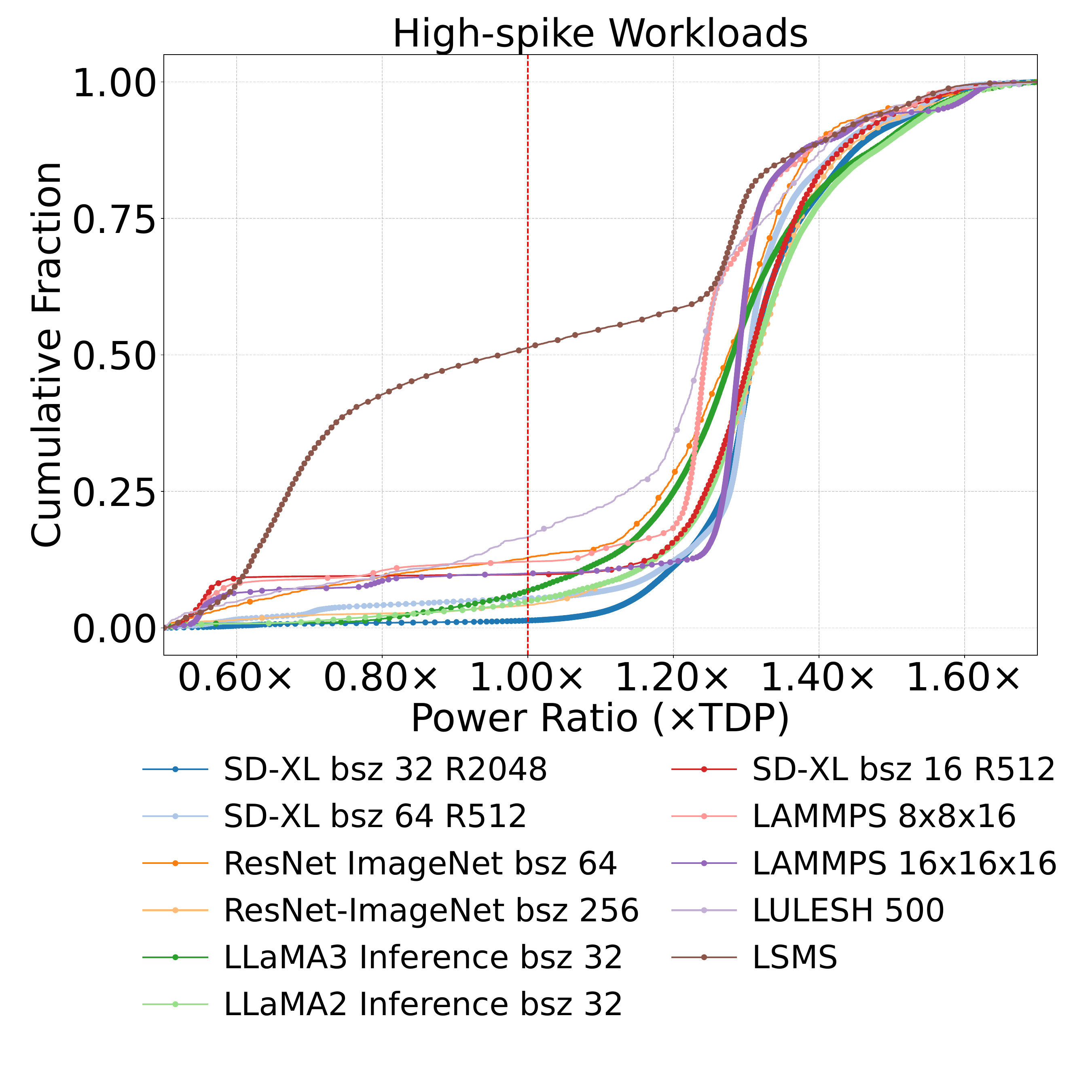}
        \caption{High Spike Workloads}
        \label{fig:high_spike}
    \end{subfigure}%
    \hspace{-0.5em}
    \begin{subfigure}[b]{0.33\textwidth}
        \centering
        \includegraphics[width=0.9\textwidth]{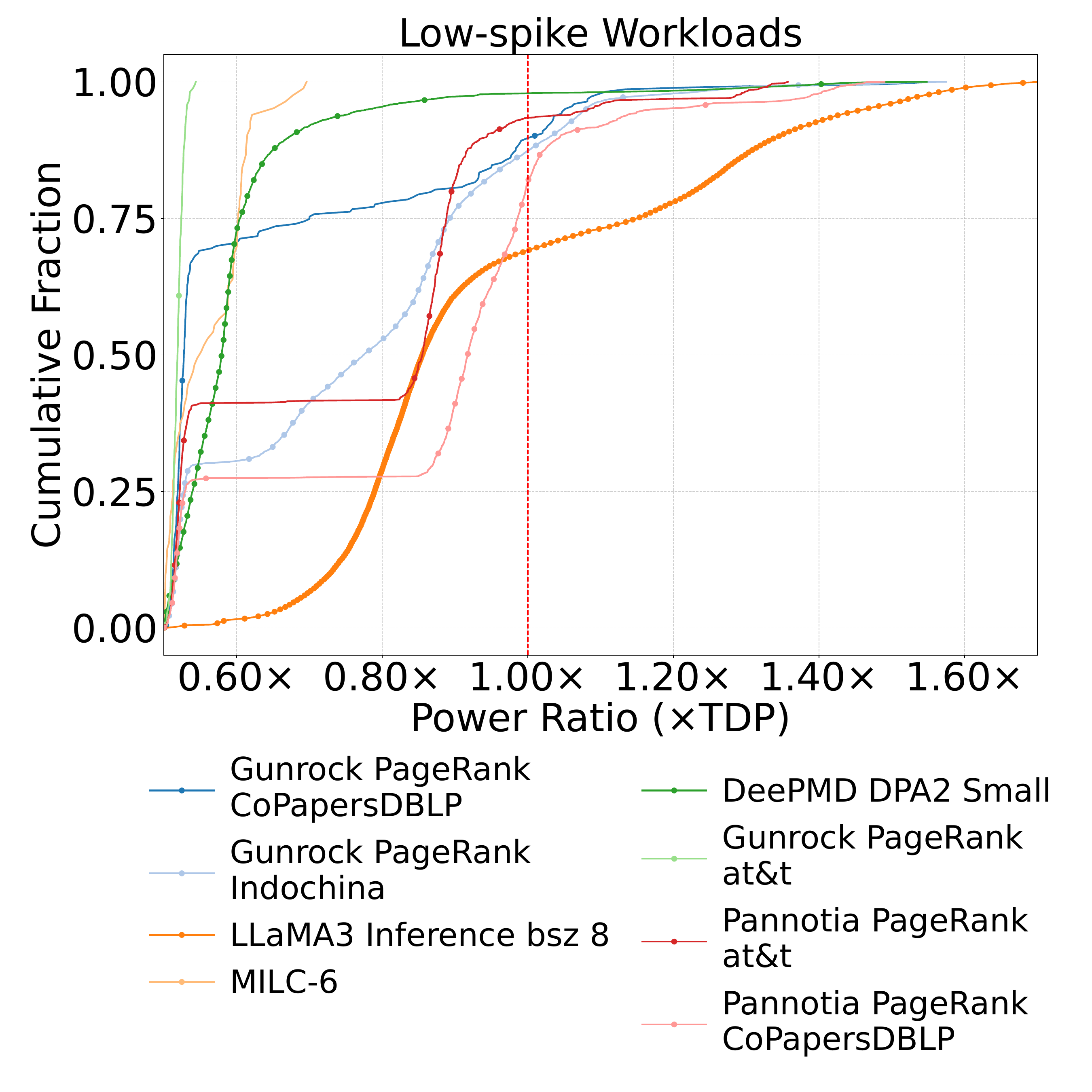}
        \caption{Low Spike Workloads}
        \label{fig:low_spike}
    \end{subfigure}%
    \hspace{-0.5em}
    \begin{subfigure}[b]{0.33\textwidth}
        \centering
        \includegraphics[width=0.9\textwidth]{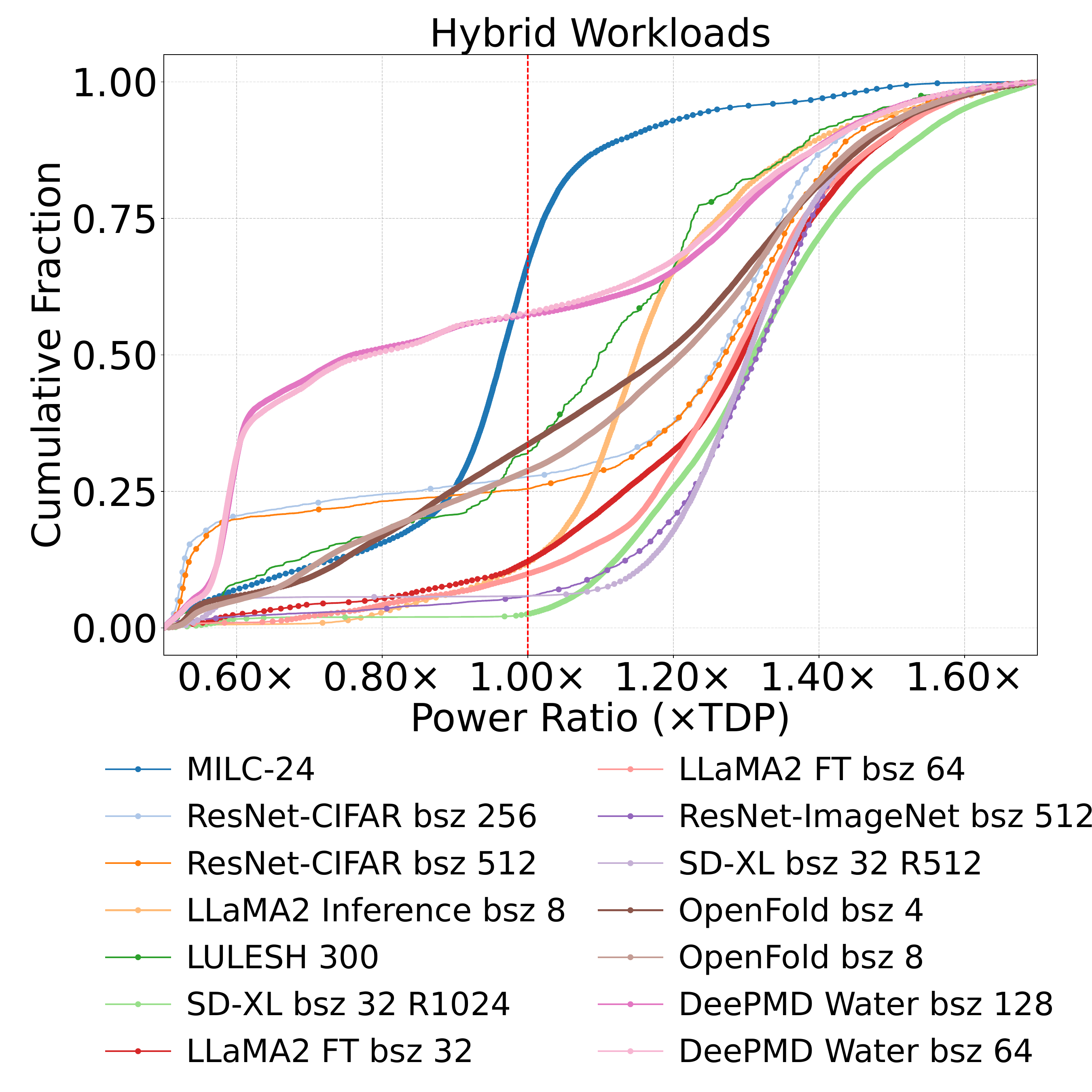}
        \caption{Mixed Workloads}
        \label{fig:hybrid}
    \end{subfigure}
    \vspace{-2ex}
    \caption{Cumulative power distributions showing power spikes for three categories of workloads}
    \Description{Power Consumption CDF per Category.} 
    \label{fig:cdf_comparison}
\end{figure}

A workload's resource utilization pattern influences its power distribution.
For example, consider the cluster of \textbf{High-spike workloads} in Figure~\ref{fig:power-cluster} that groups ML workloads like Stable Diffusion and LLaMA Inference with HPC simulations such as LSMS, LULESH, and LAMMPS. 
Figure~\ref{fig:cdf_comparison}(a) shows the cumulative power distributions for these High-spike workloads.
Other than LSMS, around 90\% of these workloads's power distributions  exceed the TDP.
Moreover, these applications exhibit a sharp vertical rise around 1.25$\times$ TDP.
Thus, a large fraction of power samples lie between 1.25$\times$ and 1.4$\times$ TDP. 
For LSMS, the upper part of the CDF (> 1.30$\times$ TDP) strongly matches the others' vertical rise, so the dendrogram groups it with the other High-spike workloads, even though 50\% of its samples are lower than TDP.
The K-Means clustering results (Figure~\ref{fig:util-cluster}) help explain these applications' power spike behavior. 
Besides LSMS, all these workloads are classified as either Compute-intensive (C) or Hybrid-intensive (H), and the compute throughput of the C-/H-class applications exceeds 40\%. 
Thus, despite different objectives and application domains, high-spike workloads share a compute-intensive signature.
For example, LAMMPS 8x8x16 (C3) and LLaMA2 Inference bsz 32 (C7) exhibit high compute throughput and relatively low memory bandwidth usage. 
Due to this compute intensity, they frequently operate in power regimes that exceed the TDP limit.
Overall, given compute power dominates 
GPU power consumption~\cite{PatelChoukse2024-POLCA}, we find a strong correlation between workloads heavily utilizing GPU compute resources and workloads with many power spikes.

Similarly, for \textbf{Low-spike workloads} (Figure~\ref{fig:cdf_comparison}(b)) the CDFs lie predominantly below the TDP limit, with over 70\% of their power samples falling well under TDP.
From a resource utilization perspective (Figure~\ref{fig:util-cluster}), many of these workloads fall into the \textit{Memory-bound (M)} category. For instance, MILC-6 (M2) and Pannotia PageRank at\&t (M3) have low compute throughput, and insufficient compute activity results in muted power profiles.

\subsubsection{Different Inputs for an Application Affect its Classification}
\label{subsubsec:eval-classes-inputs}

Figures~\ref{fig:power-cluster} and~\ref{fig:util-cluster} have several examples where different inputs for a given application have different classifications. 
For example, SD-XL bsz 64 is High-spike while bsz 32 is Mixed. 
Similarly, MILC-6 (M2), with its small problem size, falls firmly in the Low-spike and Memory-bound category.
Conversely, the larger MILC-24 (H4) is in the Hybrid/Mixed clusters for utilization and power, respectively, due to increased parallelism and balanced resource use.
Likewise, for LLaMA3 Inference, bsz 8 keeps power low, while larger batch sizes increase compute demand and thus are in the Mixed class.
Thus, 
larger inputs can increase compute or power demands, potentially shifting power- and/or utilization-classes.
This highlights the importance of a classification mechanism like \DESIGN{} -- applying optimizations for large inputs based on smaller input profiling is ineffective when the inputs fall in different classes. 

\subsubsection{Implementation Differences Affect Classification}
\label{subsubsec:eval-classes-impl}

Model implementation/algorithm variation can also affect a workload's compute and memory requirements. 
Therefore, it follows from Section~\ref{subsubsec:pwr-util-correlation} that \DESIGN{} can classify multiple implementations/algorithms for the same workload differently.
For example, Gunrock’s PageRank (C1) has high SM throughput and is categorized as Compute-intensive, whereas Pannotia’s PageRank (M3), evaluated on the same dataset, has high DRAM activity and limited compute utilization.
Thus \DESIGN{} classifies it as Memory-bound. 
This divergence underscores how software design shifts an application’s profile.
Their power spike CDFs (Figure~\ref{fig:cdf_comparison}(b)) also differ.
Pannotia's implementation has a "shelf" at lower TDP values.
This is due to the two constituent kernels \texttt{pagerank2} and \texttt{spmv\_csr\_scalar\_kernel} driving different amounts of compute -- the values above the shelf come from \texttt{spmv\_csr\_scalar\_kernel}'s power spikes.
However, despite these differences, \DESIGN{} classifies both as Low-spike since both have very few spikes over TDP. 
Thus, while Gunrock PageRank could be used to predict power distribution for the Pannotia variant (to some extent), it cannot provide performance scaling information.


Overall, these CDFs reveal correlations between workload characteristics and power dynamics, and how these vary with a workload's compute and memory balance, inputs, and 
implementation.

\begin{figure}[tb!]
    \centering
    \includegraphics[width=\linewidth]{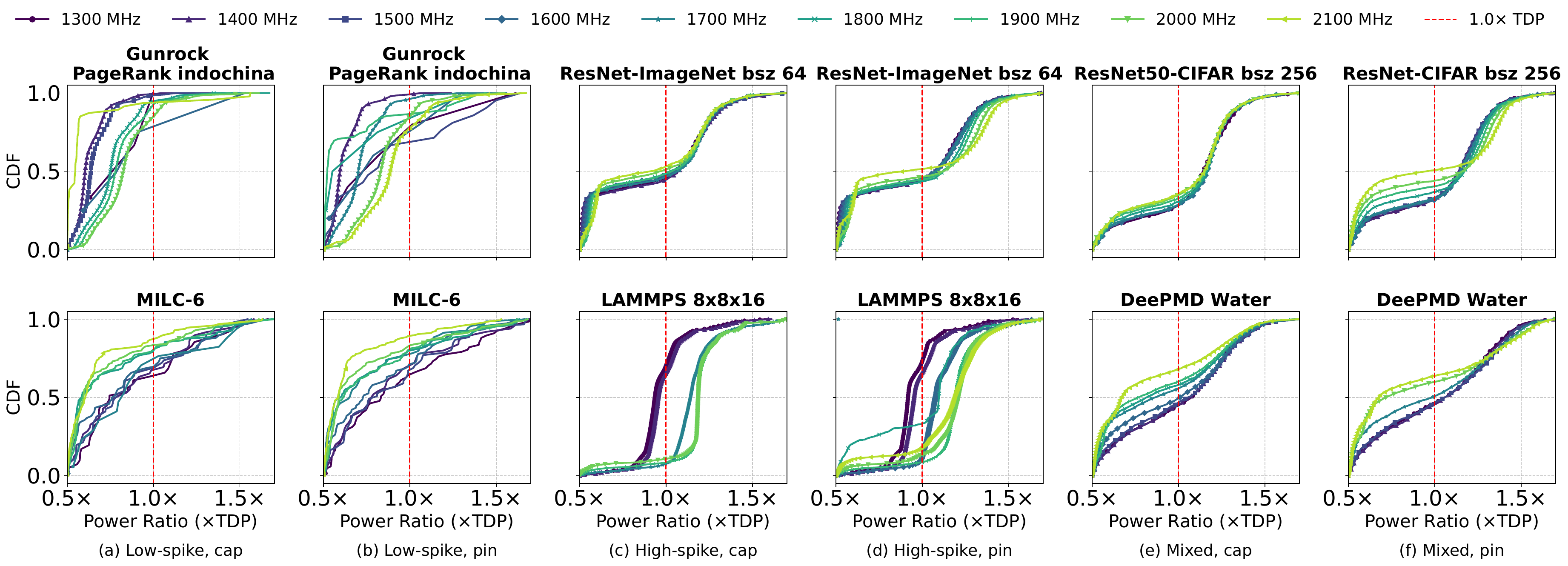}
    \caption{CDFs showing the impact of frequency capping and pinning on power spikes for different workloads grouped together by Minos in Figure~\ref{fig:power-cluster}. Each column corresponds to pairs of applications in the same class -- Low-spike, High-spike, and Mixed, under capping and pinning.}
    \Description{CDFs showing impact of frequency capping on power spikes for different workloads. Shows that workloads grouped together have similar changes in CDF.}
    \label{fig:perappfreq}
    \vspace{-2ex}
\end{figure}

\subsection{Impact of Frequency Capping/Pinning}
\label{subsec:eval-freq-cap}

Next, we evaluate \DESIGN{}'s impact on frequency capping and frequency pinning (Section~\ref{sec:back}) on the power spike distributions for a subset of the workloads in Table~\ref{tab:workloads-util}. 
Figure~\ref{fig:perappfreq} compares the changes in the CDFs as frequency is sweeping from 2100 (uncapped) to 1300 MHz, for pairs of applications grouped together in Figure~\ref{fig:power-cluster}'s dendrogram: Gunrock's PageRank \texttt{indochina} and MILC-6 from the Low-spike workloads, ResNet-ImageNet bsz 256 and LAMMPS 8x8x16 from the High-spike workloads, and DeePMD-water and ResNet-CIFAR bsz 256 from the Mixed workloads.

Since dynamic power is proportional to frequency, running at a lower frequency cap forces the GPU compute resources to run at a lower effective TDP point.
Thus, we expect the power distributions of compute-heavy applications to shift to the left as frequency capping reduces from 2100 MHz to 1300 MHz.
For applications that are memory-bound, we expect little change in the power distribution as frequency is changed. 
With frequency pinning, we expect a similar trend, but pinning can force a workload to operate at higher frequencies than it can naturally sustain, resulting in more power spikes compared to frequency capping.

Figure~\ref{fig:power-cluster} groups PageRank and MILC-6 together in the low-spike group; Figure~\ref{fig:perappfreq}'s CDFs also show that across frequency caps, the proportion of power spikes for both workloads is lower compared to other classes.
Despite having fewer power spikes, \DESIGN{} classifies Gunrock's PageRank (\texttt{indochina}) into the compute category (C4). 
This classification is validated by the left-shift trend for its CDFs in Figure~\ref{fig:perappfreq}(a) and (b), something we expect from workloads with some sensitivity to compute frequency.
For a given frequency value, the PageRank CDFs also show that frequency capping is more effective at reducing magnitudes of excursions than pinning. 
Regardless, the trends are similar across pinning and capping.
Finally, MILC-6 (M2) is a memory-bound, low-spike workload, so its CDFs do not change significantly as the CU/SM frequency is capped.
Thus, two low-spike workloads can have differing scaling behavior due to their different utilization clusters.

Figure~\ref{fig:perappfreq}(c) and (d) show the behavior of ResNet50-\texttt{ImageNet bsz 256} and LAMMPS-\texttt{8x8x16} under capping and pinning, respectively.
Both exhibit the distinct shift-left trend, as well as S-shaped CDF curves across frequencies, particularly for LAMMPS under both capping and pinning.
Given these are the furthest apart by cosine distance within the High-spike cluster (Figure~\ref{fig:power-cluster}), how steep their vertical rise is differs between LAMMPS and ResNet.
Thus, it is important for \DESIGN{}'s algorithm (Section~\ref{sec:design}) to select the nearest neighbor application when predicting power distributions, not just any member of the same class. 

Finally, we examine 
two Mixed class applications, DeePMD-Water and ResNet50-CIFAR with a batch size of 256.
For these workloads, the scaling of power with frequency is in between the low- and high-spike groups. 
Instead of shifting left, both capping and pinning shift the CDF downward: the fraction of power spikes above TDP increases while the magnitude of spikes decreases.
This 
is most pronounced for ResNet-CIFAR bsz 256 under pinning (Figure~\ref{fig:perappfreq}(f)), where the proportion of power samples above TDP increases from 40\% at 2100 MHz to 68\% at 1300 MHz.
This happens because capping effectively reduces power samples that lie below TDP, but cannot reduce spikes over the limit. 
As a result, the workload's fraction of spikes over TDP increases. 
These shared characteristics show it is important to distinguish these workloads from the High-spike category to accurately predict their power consumption patterns under frequency limits.

\begin{figure}[tb!]
    \begin{subfigure}{0.325\textwidth}
        \includegraphics[width=\linewidth]{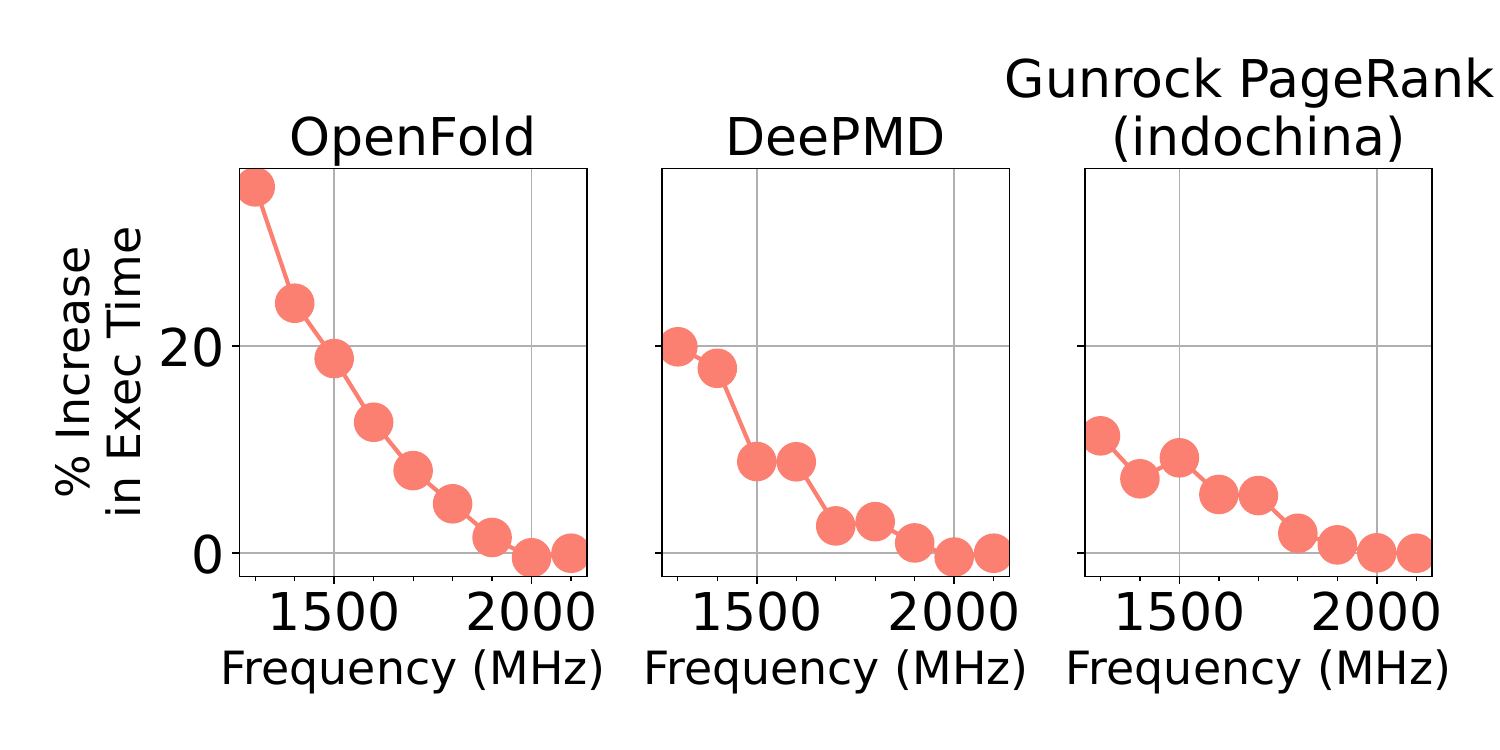} 
        \caption{Compute-intensive workloads.}
        \Description{Compute-intensive workloads.}
        \label{fig:per-scaling-c}
    \end{subfigure}
    \begin{subfigure}{0.325\textwidth}
        \includegraphics[width=\linewidth]{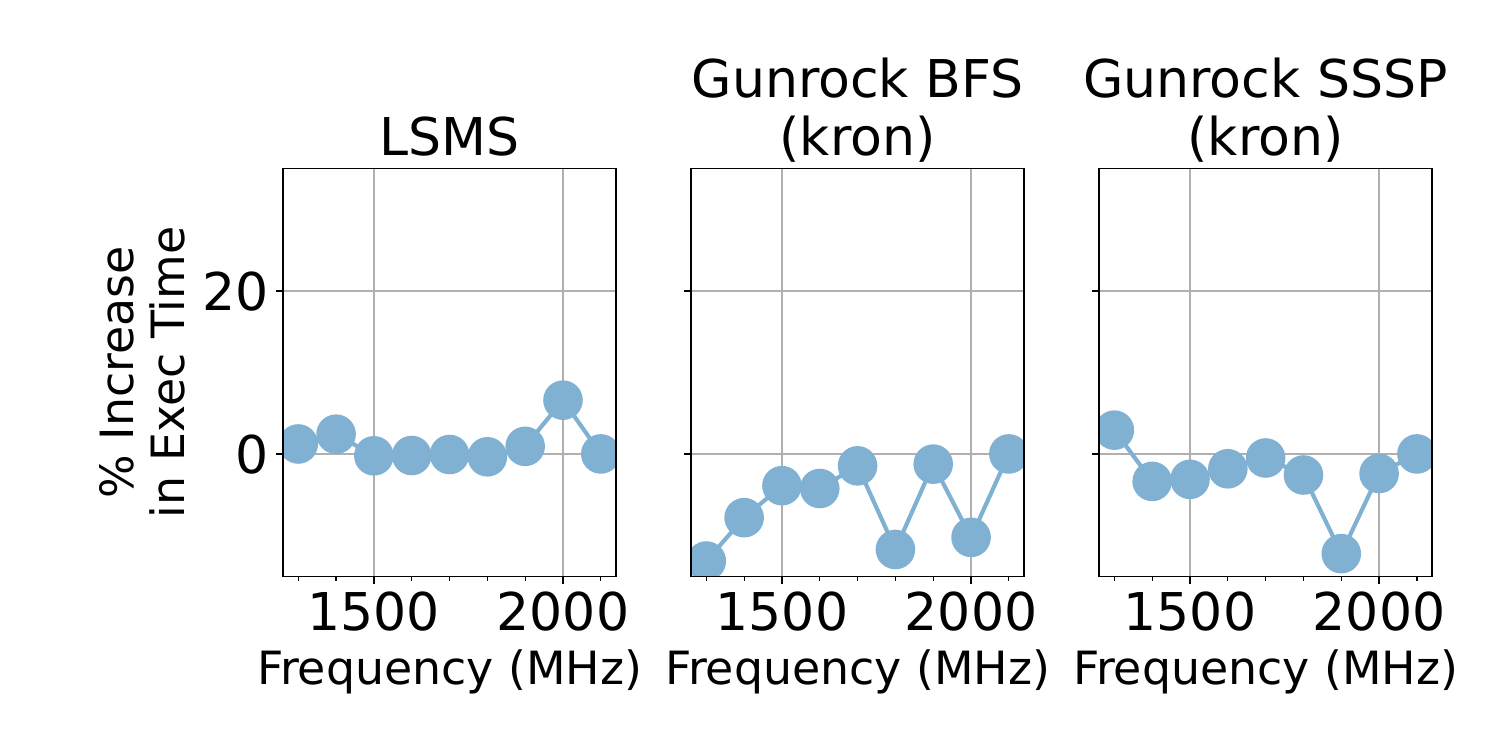}
        \caption{Memory-intensive workloads.}
        \Description{Memory-intensive workloads.}
        \label{fig:per-scaling-mem}
    \end{subfigure}
    \begin{subfigure}{0.325\textwidth}
        \includegraphics[width=\linewidth]{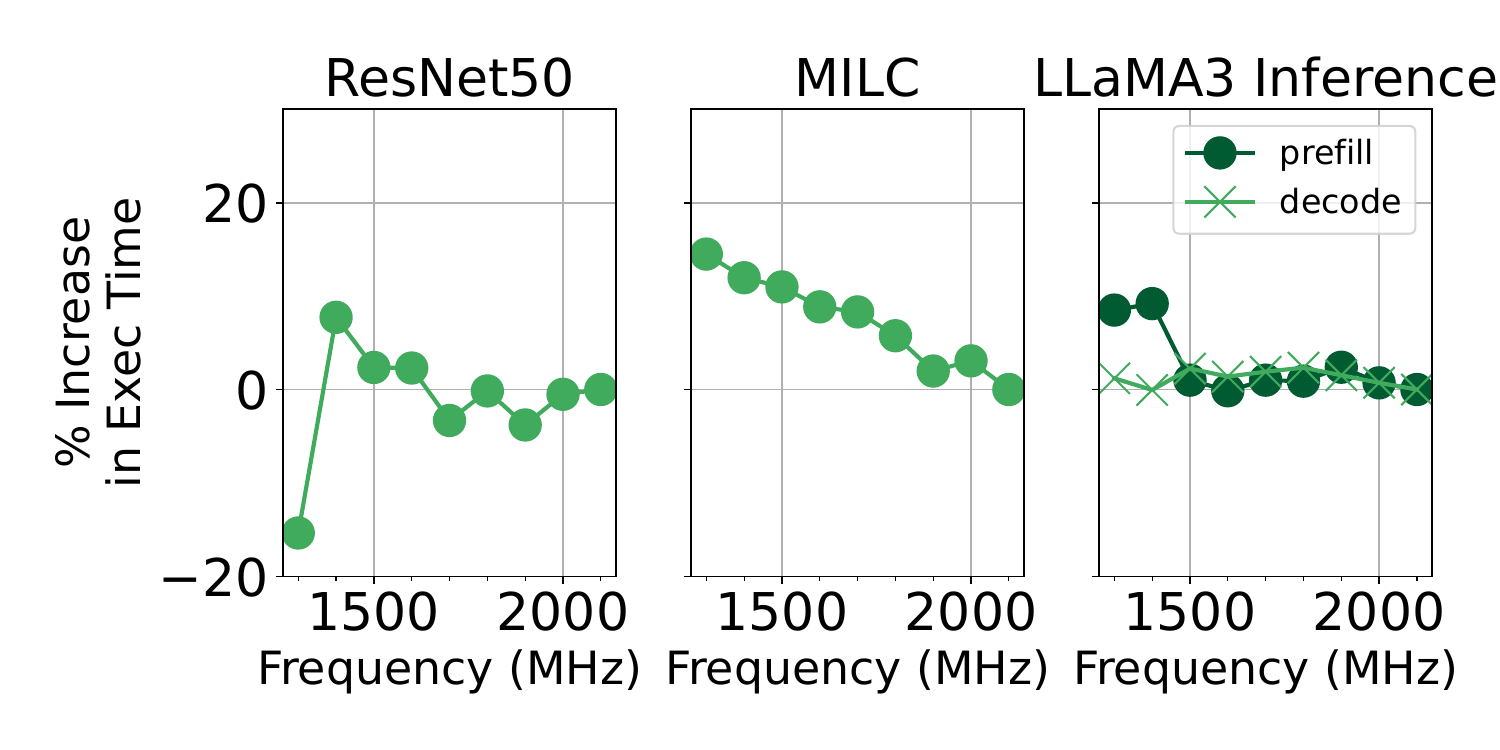}
        \caption{Hybrid workloads.}
        \Description{Hybrid workloads.}
        \label{fig:per-scaling-hybrid}
    \end{subfigure}
    \vspace{-3ex}
    \caption{Performance scaling with frequency for C-, H-, and M-class applications. Execution time increase at each frequency cap is relative to uncapped runs (2100 MHz).} 
    \Description{Performance scaling with frequency for C- and M-class applications as per our utilization-based classification.}
    \label{fig:perf-freq-scaling}
    \vspace{-4ex}
\end{figure}


\noindent
\textbf{Impact on Performance}: 
While frequency capping and pinning reduce a workload's power consumption and allow power oversubscription in large-scale clusters, they may also hurt performance if a workload is sensitive to SM/CU frequency.
Thus, cluster operators must also consider this.
\DESIGN{}'s utilization-based classification provides a mechanism to estimate a workload's performance scaling under frequency limits.
Figure~\ref{fig:perf-freq-scaling} shows the percentage increase in execution time (performance degradation) for a subset of workloads as frequency caps vary from 1300 MHz to uncapped (2100 MHz).
Unsurprisingly, compute-intensive applications (e.g., DeePMD (C9), PageRank (C4), and OpenFold (C2)) are highly sensitive to capping.
Their performance degrades by 34\%, 11\%  and 20\%, respectively, when capping frequency to 1300 MHz versus the uncapped 2100 MHz baseline.

Conversely, memory-intensive applications like Breadth First Search (BFS), Single-Source Shortest Path (SSSP), and LSMS show little to no performance variation when capping compute frequency to lower values, validating our memory-bound classification.

The hybrid workloads have mixed behavior.
ResNet50 (H2) has up to a 10\% absolute difference in training iteration times, while MILC-24 (H4) gracefully degrades performance by $\approx$14\%.
For LLaMA3 inference (H1), we split the performance into two SLOs: the prefill phase latency or Time To First Token (TTFT) and per-token decode latency or Time Between Tokens (TBT).
Capping frequencies to lower values hurts performance for the compute-intensive prefill phase,
while the memory-bound decode phase is largely unaffected by frequency caps.
This separation explains why MILC-24 and ResNet50 do not show a consistent pattern: some of their kernels are compute-intensive while others are memory-intensive.
Thus, their response to scaling frequency is impacted by this mix.


\section{Evaluation}
\label{sec:eval-usage}
In this section, we show that \DESIGN{} accurately predicts the optimal frequency capping configuration for workloads not in its reference set.
Specifically, we show that \DESIGN{} selects FAISS's and Qwen1.5-MoE's optimal frequency caps for power spike predictions with 5\% error and performance predictions with 0\% prediction error, while reducing profiling time by 89-90\%. 
When extending this to other workloads (Section~\ref{subsec:eval-pairwise}) \DESIGN{} achieves 4\% average error when predicting 90th percentile (p90) power spikes and 3\% average error when predicting performance loss.
\DESIGN{} also reduces the average prediction error from 14\% to 4\% versus Guerreiro, et al.'s state-of-the-art classification scheme~\cite{Guerreiro2019dvfs} (Section~\ref{subsec:eval-guerreiro}).
Thus, \DESIGN{} demonstrates the importance of balancing performance, power, and power spikes when classifying workloads.

\subsection{Case Study: FAISS and Qwen1.5-MoE}
\label{subsec:eval-case-study}
\Edit{We further demonstrate \DESIGN{}'s effectiveness by determining optimal frequency caps for a never-before-seen workload which is not a part of the applications used in Section~\ref{subsec:eval-classes}.
We use Algorithm~\ref{alg:minos} to find a frequency cap such that we can ensure safe operating limits without having to perform extensive power distribution profiling at different frequencies. 
As discussed in Section~\ref{subsec:design-usingminos}, determining what frequency cap to run a workload with involves managing power and performance tradeoffs.
To show how \DESIGN{} balances these, consider two new applications which we have no prior information about: \textbf{FAISS}, a vector similarity search workload~\cite{faiss-gpu} and \textbf{Qwen1.5-MoE-A2.7B}, a Mixture-of-Experts (MoE) transformer-based inference model~\cite{qwen1.5-hf}.
Unless otherwise stated, we use \textbf{Qwen1.5-MoE} to abbreviate Qwen1.5-MoE-A2.7B throughout. 
\Comment{We chose \textbf{FAISS} specifically because it performs large-scale nearest neighbor searches via batched matrix-vector distance computations, a significantly different workload pattern from the workloads in our reference set. 
Similarly, we chose \textbf{Qwen1.5-MoE} to predict optimal settings for an MoE, whereas the reference set contains decoder-only LLMs.}
We only collect power and performance counter profiles at the default, uncapped frequency as input to \DESIGN{}'s classification. 
Following Algorithm~\ref{alg:minos}, we identify the nearest performance and power neighbors for these workloads, and use their frequency scaling data to predict optimal frequency settings for these workloads. 
Table~\ref{tab:case-study} lists the power and performance neighbors ($R_{pwr}$ and $R_{perf}$) for these new applications along with their distances to the target workloads.
For \textbf{FAISS}, both $R_{pwr}$ and $R_{perf}$ are \textbf{SD-XL}, whereas for \textbf{Qwen1.5-MoE}, $R_{pwr}$ is \textbf{MILC-24} and $R_{perf}$ is \textbf{DeePMD Water}.}

\Edit{Next we use the respective neighbors' profiling data to determine optimal frequency caps for these new applications (Algorithm~\ref{alg:minos}). 
Then we run FAISS and Qwen1.5-MoE and evaluate how well our predictions align with these workloads' real behavior.
As mentioned in Section~\ref{subsec:design-usingminos}, frequency capping can be done for different objectives.
We consider both scenarios: (1) if workload SLOs allow for some slack in performance (\textsc{PowerCentric}), keep all excursions strictly under a threshold and (2) minimize power spikes while keeping performance strictly bounded (\textsc{PerfCentric}).}

{ \footnotesize
  \centering
    \begin{table}[tb!]
    \centering
    \vspace{1ex}
    \caption{New applications and their nearest neighbors according to \DESIGN' classification}
    \label{tab:case-study}
    \vspace{-1ex}
    \begin{tabular}{@{}lllll@{}}
    \toprule
    \multicolumn{1}{c}{New Application} & \multicolumn{1}{c}{\begin{tabular}[c]{@{}c@{}}Power Neighbor\\  ($R_{pwr}$)\end{tabular}} & \multicolumn{1}{c}{\begin{tabular}[c]{@{}c@{}}Cosine Distance \\ to $R_{pwr}$\end{tabular}} & \multicolumn{1}{c}{\begin{tabular}[c]{@{}c@{}}Perf Neighbor \\ ($R_{perf}$)\end{tabular}} & \multicolumn{1}{c}{\begin{tabular}[c]{@{}c@{}}Euclidean Distance \\ to $R_{perf}$\end{tabular}} \\ \midrule
    FAISS bsz 4096  & SD-XL                      & 0.05                         & SD-XL                      & 7.18                             \\
    Qwen1.5B bsz 32 & MILC-24             & 0.01                         & DeePMD Water              & 13.64                            \\ \bottomrule
    \end{tabular}
    \end{table}
}
\begin{figure}[tb!]
    \centering
    \includegraphics[width=\linewidth]{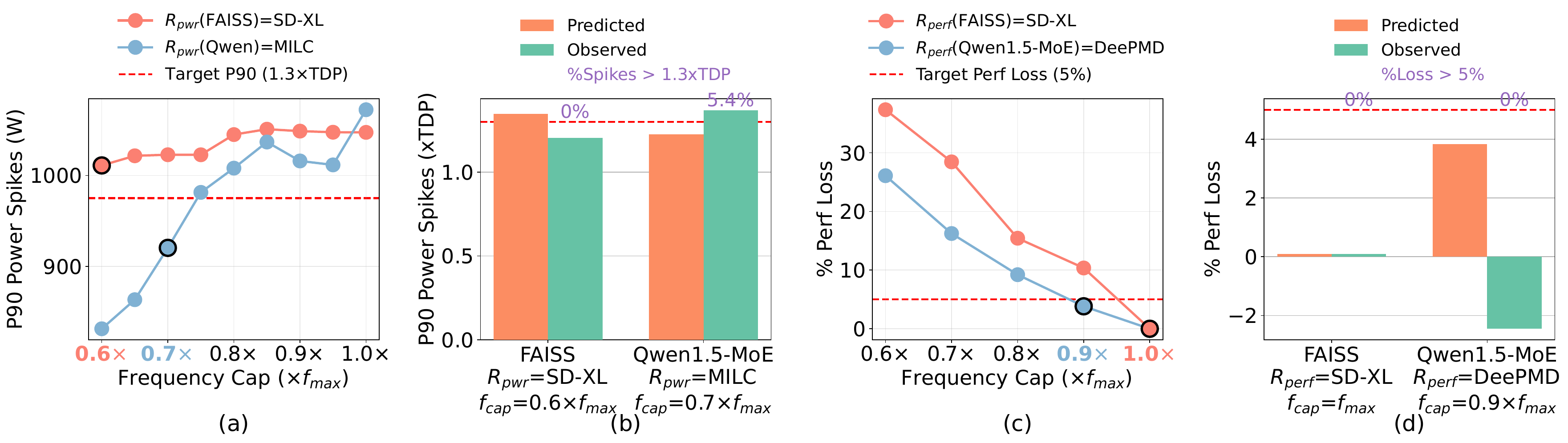}
    \vspace{-3ex}
    \caption{\Edit{(a) p90 power spike scaling with frequency caps for power neighbors, (b) p90 power prediction errors for FAISS and Qwen1.5-MoE, (c) performance scaling with frequency caps for performance neighbors, and (d) performance prediction errors for FAISS and Qwen1.5-MoE.}  \Comment{In (b), prediction error$=(\text{observed p90 power} - 1.3 \times \text{TDP}) / 1.3 \times \text{TDP}$: positive values indicate the p90 spikes exceeded the $1.3\times\text{TDP}$ target(e.g., 5.4\% for Qwen1.5-MoE), 0\% means at or below target. In (d), prediction error$=(\text{observed perf loss} - 5\%)$: 0\% means within the 5\% bound}}
    \Description{(a) p90 power spike scaling with frequency caps for power neighbors, (b) p90 power prediction errors for FAISS and Qwen1.5-MoE, (c) performance scaling with frequency caps for performance neighbors, and (d) performance prediction errors for FAISS and Qwen1.5-MoE.}
    \label{fig:case-study-errors}
    \vspace{-1ex}
\end{figure}

\subsubsection{\textsc{PowerCentric} Approach}
\label{subsubsec:eval-case-study-power}
\Edit{First, we select a frequency cap that can strictly bound the 90th percentile of power spikes, while some degradation in performance is acceptable.}
Such optimizations are common when clusters are over-provisioned for power and must aggressively frequency-cap workloads to maintain a system-level power budget~\cite{PatelChoukse2024-POLCA, Patel2024Splitwise, Stijkovic25TAPAS}.
We assume p90 power spikes to be at or below $1.3\times$ TDP.
We use the power spike scaling of FAISS and Qwen1.5-MoE's power neighbors, namely SD-XL and MILC, to identify the optimal frequency cap for these workloads.
Figure~\ref{fig:case-study-errors}(a) shows how the 90th percentile spikes scale with different caps for the neighbor workloads ($1.3\times$TDP threshold shown as a red dashed line). 
The p90 spikes cross the threshold at $0.6\times f_{max} =$ 1300 MHz for SD-XL and $0.7\times f_{max} =$ 1500 MHz for MILC. 
\DESIGN{} predicts these optimal frequency caps would limit the new workloads' p90 power spikes to $1.3\times$TDP.
\Comment{While we use p90 power as an example, system administrators may alternatively want to constrain 95th percentile (p95), 99th percentile (p99), or peak power, depending on provisioned system power budgets and GPU power tolerances. 
In Section~\ref{sec:pairwise-pwr} we show how \DESIGN{}'s accuracy changes as we constrain for p95 or p99 power.}

Figure~\ref{fig:case-study-errors}(b) compares predicted p90 spikes with the workload's actual observed spikes.
SD-XL is a perfect predictor for FAISS, with p90 spikes being strictly limited below the threshold. 
Although MILC slightly under-predicts Qwen1.5-MoE's p90 spikes, the prediction error is only 5\%. 
Thus, \DESIGN{} effectively predicts optimal frequency caps without requiring detailed profiling of new workloads at different frequency caps.

\subsubsection{\textsc{PerfCentric} Approach}
\label{subsubsec:eval-case-study-perf}
\Edit{Next, we set an optimal frequency cap for FAISS and Qwen1.5-MoE that strictly limits performance degradation to within $<$ 5\% (the same target as POLCA~\cite{PatelChoukse2024-POLCA}). 
Table~\ref{tab:case-study} shows the performance neighbors for FAISS and Qwen1.5-MoE are SD-XL and DeePMD (C9), respectively. 
Figure~\ref{fig:case-study-errors}(c) shows SD-XL and DeePMD's performance degradation is linear with frequency caps, with the 5\% performance loss threshold marked as a red dashed line.
Here, we want to select the lowest possible frequency that gives us $<$ 5\% degradation and if possible reduces power spikes. 
Thus, our new workload's optimal cap is the frequency at which the neighbor's performance loss is strictly below 5\%: $0.9\times f_{max}$ for Qwen1.5-MoE's neighbor DeePMD and uncapped for FAISS's neighbor SD-XL.
Figure~\ref{fig:case-study-errors}(d) shows applying these caps successfully keeps both FAISS and Qwen1.5-MoE's performance loss within 5\%, with 0\% prediction error for both workloads.}

\subsubsection{Profiling Savings}
\label{sec:eval-savings}
Minos requires a single profiling run at the default GPU clock frequency, and predicts performance and power scaling across all frequencies by matching it to its nearest neighbor application.
Compared to sweeping over the frequency range $f \in \mathcal{F}$, this reduces profiling time by: $ 1-(T_{f0}/\sum_{f\in \mathcal{F}}T_f)$. For the two case study workloads, Minos saves \textbf{90}\% profiling time for FAISS and \textbf{89}\% for Qwen1.5-MoE while providing accurate frequency scaling predictions.


\subsection{Generalization Across Workloads}
\label{subsec:eval-pairwise}

\begin{figure}[tb!]
    \centering\includegraphics[width=\linewidth]{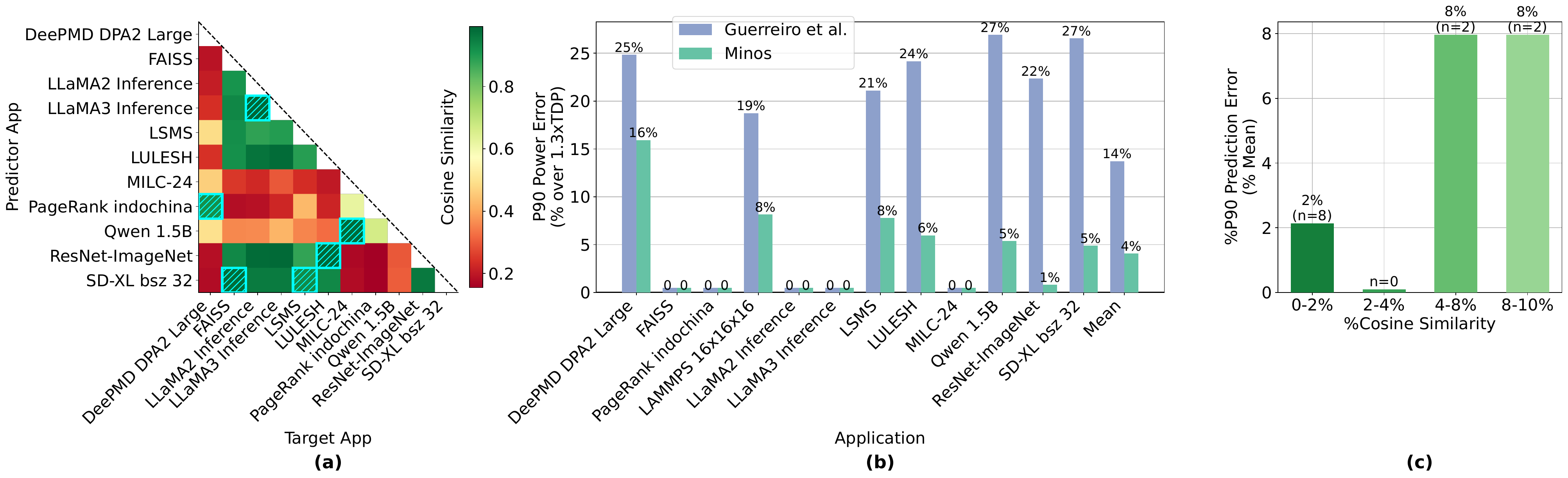}
    \vspace{-4ex}
    \caption{\Edit{(a) Pairwise matrix showing cosine similarity between workloads and nearest power neighbors highlighted in \textcolor{cyan}{cyan}, (b) p90 power prediction errors using Guerreiro et al.~\cite{Guerreiro2019dvfs} and \DESIGN{}, and (c) histogram of \DESIGN{}'s p90 prediction errors across workload pairs, binned by cosine distance between workloads.}}
    \Description{(a) Pairwise similarity matrix showing nearest power neighbors for each application, (b) p90 power prediction errors using power neighbor scaling for Guerreiro et al.~\cite{Guerreiro2019dvfs} and \DESIGN{}, and (c) Histogram of p90 power prediction errors across workload pairs, binned by percentage of cosine distance between workloads.}
    \label{fig:pwr-pairwise}
    \vspace{-2ex}
\end{figure}

\Edit{To determine the generality of \DESIGN{}'s frequency cap selection, we adopt 
a \textbf{hold-one-out cross-validation}~\cite{Arlot_2010}.
Hold-one-out cross-validation is widely used to assess a model’s ability to generalize to unseen data.
To avoid different inputs of the same workload being neighbors, we only consider one input per workload.
Specifically, we consider the largest batch size or input for each workload in Table~\ref{tab:workloads-util}.
Treating every unique application as a new workload, we identify its nearest power and performance neighbors (Algorithm~\ref{alg:minos}).
Then, \DESIGN{} uses the power and performance scaling information from the neighbor to predict the held-out application's behavior. 
Similar to Section~\ref{subsec:eval-case-study}, we examine both power- and performance-centric approaches.}

\subsubsection{\textsc{PowerCentric} Approach}
\label{sec:pairwise-pwr}
\Edit{Figure~\ref{fig:pwr-pairwise}(a) shows the pairwise-similarity between unique applications as a triangular heatmap matrix. 
This identifies the nearest power neighbor for each workload and how close the workloads' power spike distributions are to their neighbor. 
Figure~\ref{fig:pwr-pairwise}(b) further shows the prediction errors in p90 power spikes for each workload based on its neighbor.
Overall, \DESIGN{} successfully predicts optimal frequency caps to limit power spikes within the specified threshold, with 4\% error on average across workloads.
While workloads like FAISS and SD-XL (Section~\ref{subsec:eval-case-study}) show high cosine similarity and therefore 0\% prediction errors, other pairs such as PageRank indochina and DeePMD DPA2 Large diverge in their power spike distributions by up to 10\%. 
Distributing prediction errors across cosine similarity bins  (Figure~\ref{fig:pwr-pairwise}(c)), prediction errors grow larger as the cosine distance between the workload and its power neighbor increases. 
Thus, if the cosine distance to the nearest workload is greater than 10\%, \DESIGN{}'s workload space needs more points or workloads to make more accurate predictions.
Thus for any new workload, the cosine distance to its neighbor can determine the expected accuracy of \DESIGN{}'s prediction results.} 

\begin{figure}[tb!]
    \centering
    \includegraphics[width=0.4\linewidth]{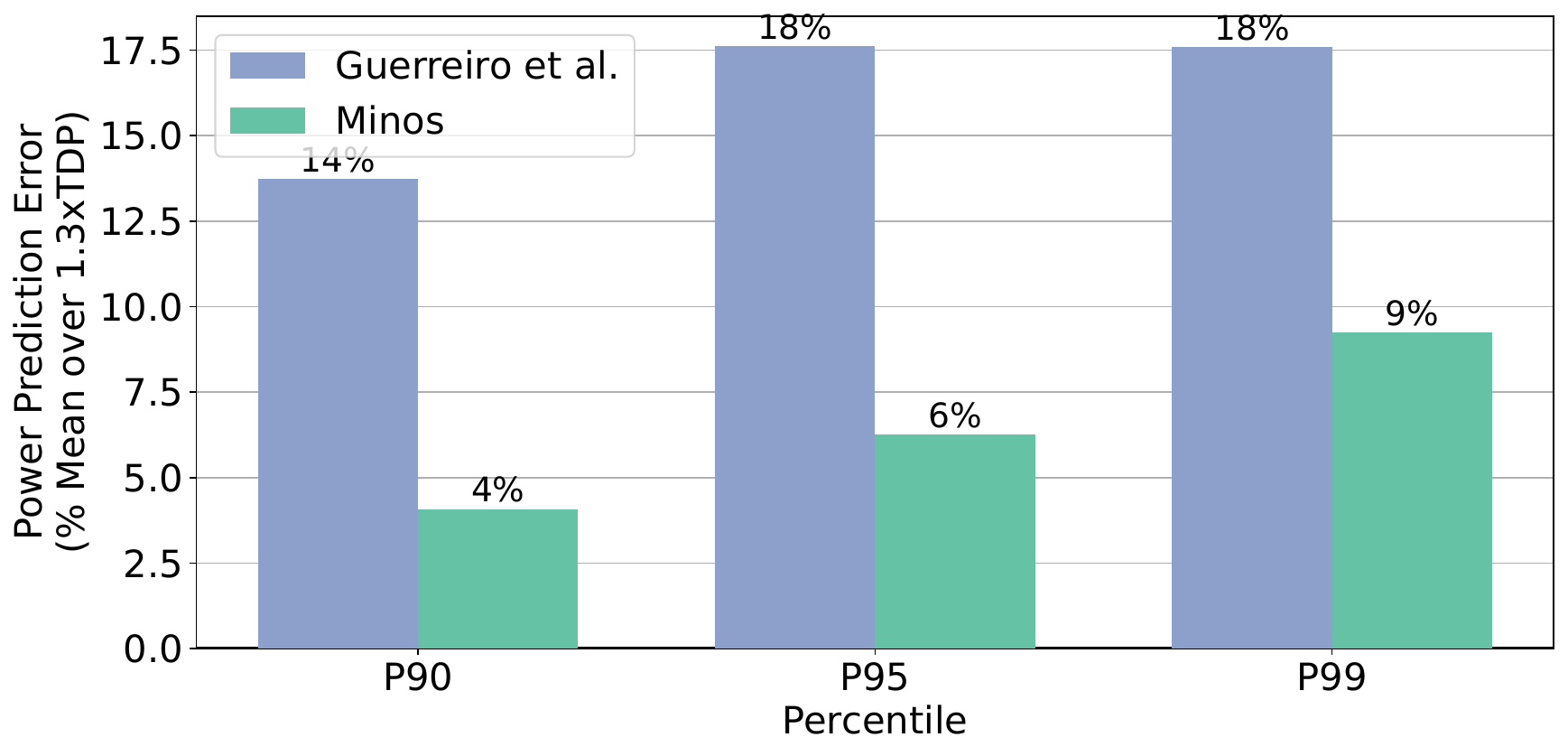}
    \vspace{-2ex}
    \caption{\Comment{Comparing the p90, p95, and p99 power prediction errors for Guerreiro et al.~\cite{Guerreiro2019dvfs} versus \DESIGN{}.}}
    \Description{Comparing the p90, p95, and p99 power prediction errors for Guerreiro et al.~\cite{Guerreiro2019dvfs} versus \DESIGN{}.}
    \label{fig:p95-p99-agg}
    \vspace{-2ex}
\end{figure}

\Comment{When setting frequency caps for a \textsc{PowerCentric} experiment, users might want to constrain p95 or p99 power spikes instead of p90 spikes.
Accordingly, Figure~\ref{fig:p95-p99-agg} shows the average prediction errors when the 95th and 99th percentile of power spikes are constrained to $1.3\times$TDP. 
Overall, when constraining the system for higher power spike percentiles, the average prediction error slightly increases from 4\% with p90 to 6\% with p95 and 9\% for p99.
Most workloads are relatively unaffected by this change.
However a few workloads with worse cosine distances, such as DeePMD DPA2 Large, are impacted -- increasing the average prediction error.
Regardless, \DESIGN{} is still able to accurately predict power spikes within 9\% average error across workloads and different power spike thresholds.}

\begin{figure}[tb!]
    \centering\includegraphics[width=\linewidth]{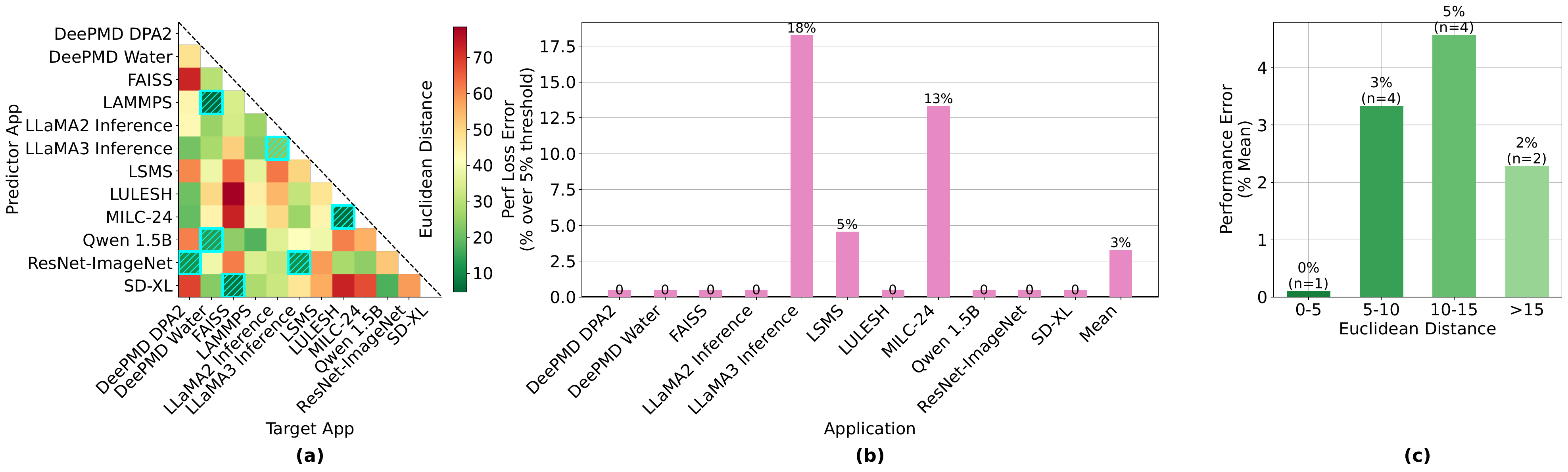}
    \vspace{-4ex}
    \caption{\Edit{(a) Matrix showing euclidean distance between workload pairs with nearest neighbors $R_{perf}$ highlighted in \textcolor{cyan}{cyan}, (b) performance scaling prediction errors with \DESIGN{}, and (c) histogram of performance prediction errors across workload pairs, binned by euclidean distance between workloads.}}
    \Description{(a) Pairwise similarity matrix showing nearest neighbors $R_{perf}$ for each application, (b) performance scaling prediction errors with \DESIGN{}' PerfCentric algorithm, and (c) Histogram of performance prediction errors across workload pairs, binned by percentage of euclidean distance between workloads.}
    \label{fig:perf-pairwise}
    \vspace{-3ex}
\end{figure}

\subsubsection{\textsc{PerfCentric} Approach}
\label{sec:pairwise-perf}
\Edit{Next, we generalize Section~\ref{subsubsec:eval-case-study-perf}'s performance-centric experiments by considering nearest performance neighbors $R_{perf}$ for all applications.
Figure~\ref{fig:perf-pairwise}(a) shows them, with the heatmap color indicating euclidean distance between them. 
Figure~\ref{fig:pwr-pairwise}(b) shows the prediction errors in performance degradation across workloads: 3\% average prediction error, and perfect predictions for 8 out of 11 unique workloads.
Finally, Figure~\ref{fig:perf-pairwise}(c) distributes these errors into bins of euclidean distance ranges. 
While the average error increases as euclidean distance between workloads increases, the trend is not as well-defined as Figure~\ref{fig:pwr-pairwise}(c). 
However, in practice one can impose a minimum allowable frequency, since extremely low predicted caps would severely degrade performance and are impractical in real deployments.
Adding this lower bound eliminates the low-frequency outliers while still following \DESIGN{}'s algorithm to predict performance scaling.}

\subsection{Comparison with State-of-the-art}
\label{subsec:eval-guerreiro}

\Edit{We also compare \DESIGN{} against Guerreiro et al.~\cite{Guerreiro2019dvfs}, the most relevant prior work, which uses mean power in their classification scheme (Sections~\ref{sec:back} and~\ref{sec:related}).
Figure~\ref{fig:pwr-pairwise}(b) shows power prediction errors using their methodology in addition to the errors with \DESIGN{}.  
Overall, \DESIGN{} (4\% average error) significantly reduces prediction errors versus Guerreiro's approach (14\%).
For low-spike such as PageRank indochina and some hybrid-spike workloads, using mean power is sufficient to make accurate predictions. 
For these workloads, power spikes are below the specified threshold for a large range of frequencies.
Thus, both methodologies make highly accurate predictions. 
However, workloads with higher spikes and/or dynamically varying power consumption behavior like DeePMD and ResNet cannot be characterized by a single mean power.
Accordingly, the Guerreiro methodology results in significantly higher prediction errors for them (Figure~\ref{fig:pwr-pairwise}(b)).
This highlights the importance of \DESIGN{}' clustering based on each workload's power spike distributions.}

\subsection{Sensitivity Analysis}
\label{subsec:eval-sensitivity}

\Edit{Bin sizes selection should balance identifying broad similarity in CDF shapes for workloads with retention of fine-grained distribution features (Section~\ref{subsubsec:meth-power-class}). 
Thus, we also evaluated the sensitivity of our classification to the choice of bin size used to generate power spike distribution vectors for each workload.
Specifically, we evaluated \DESIGN{}'
p90 power prediction errors for each application $T$ across different bin sizes $c$: $\mathrm{Err}_c(T) =\left|p90(T)-p90\!\left(\mathrm{NN}_c(T)\right)\right|$, where $p90(\cdot)$ denotes the 90th percentile of power, and $\mathrm{NN}_c(T)$ is the nearest power neighbor of $T$ with bin size $c$.
}

\begin{figure}
  \centering
  \includegraphics[width=0.5\linewidth]{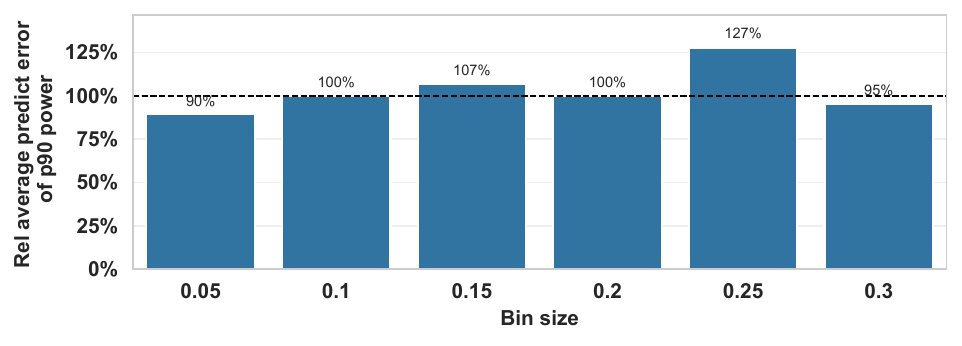}
  \vspace{-2ex}
  \caption{\Edit{Average errors for p90 power predictions with different bin sizes, normalized to bin size 0.1.}}
  \Description{Average errors for p90 power predictions with different bin sizes, normalized to bin size 0.1.}
  \label{fig:bin-size}
  \vspace{-2ex}
\end{figure}

\Edit{Figure~\ref{fig:bin-size} shows that \textbf{medium bin sizes (0.1, 0.15, and 0.2)} yield nearly identical accuracy: their average normalized errors differ by $<$10\%.
Conversely, some very large bins exhibit higher prediction errors, since they aggregate power information and lose feature richness. 
This indicates that \DESIGN{}' classification is robust to the choice of bin size if moderate granularity is used.
More broadly, Algorithm~\ref{alg:minos} overcomes this by picking the best bin size (\texttt{ChooseBinSize}).
}

\section{Discussion}
\label{sec:disc}

\noindent
\textbf{GPU Vendor Impact}: 
\DESIGN{} is applicable to any GPU vendor, as long as the vendor provides performance (e.g., CU and DRAM utilization, Section~\ref{subsec:design-util}) and power (e.g., power spikes, Section~\ref{subsec:design-pwr}) profiling information.
However, GPU vendors do not always calculate profiling information identically.
For example, AMD's and NVIDIA's profilers currently do not calculate CU utilization the same, making direct CU utilization comparisons across the vendors difficult.
Nevertheless, as long as \DESIGN{} compares profiling information relative to other profiling information from the same vendor, it can obtain similar categorization results to ours in Section~\ref{sec:eval-usage} for any GPU vendor.
Moreover, even if \DESIGN{} identifies different performance-power tradeoffs across GPU vendors, developers can still utilize these findings to optimize workloads on their specific systems.

\noindent
\textbf{Impact of GPU Generations on Power Spikes}: 
Recently, GPU vendors have been increasing computational resources (e.g., GPU FLOPS increased 7$\times$ between 2022 and 2025~\cite{AMD_MI210,AMD_MI355X}) and TDPs (Section~\ref{sec:intro}) every generation.
Consequently, workloads 
that experience power spikes on MI210- or A100-class GPUs may have different power spikes on higher TDP MI350- and B100-class GPUs.
For example, LLaMA3 Inference on an MI210 GPU (not shown due to space constraints) has power spikes up to 1.4$\times$ TDP (450W spikes, 300W TDP) during its compute-heavy prefill phase, whereas on an MI300X GPU Figure~\ref{fig:pwr-timeseries} shows its power spikes up to 1.7$\times$ TDP (1290W spikes, 750W TDP).
Moreover, GPU workloads are voracious in their computing demands~\cite{GholamiYao2024-MLGrowth, ShehabiNewkirk2024-doeDCEnergy}.
Thus, as GPU resources scale, workloads' computational needs also increase -- and future HPC cluster workloads will likely have power spikes.
However, even if new workloads exhibit different power spike behavior, \DESIGN{} is agnostic of any specific GPU vendor or generation, and Section~\ref{sec:eval-classes} shows \DESIGN{} effectively classifies the power spike behavior of diverse workloads and inputs.
As a result, \DESIGN{}' approach will effectively categorize and enable studying power spike tradeoffs for newer GPUs.


\section{Conclusion}
\label{sec:conc}

Widely used, accelerator-rich HPC systems have diverse use cases.
However, modern accelerators's power limits, workloads' requirements, and cluster sizes are all ravenously increasing.
Consequently, this causes frequent power spikes that temporarily exceed the GPU's TDP.
Collectively, these challenges make it difficult 
to co-optimize workloads for performance, power, and power spikes.
Accordingly, we design \DESIGN{}, a novel classification scheme that profiles GPU performance, power, and power spike behavior.
Across 18 modern graph analytics, HPC, HPC+ML, and ML workloads, 
\DESIGN{} successfully classifies these workloads's behavior, and its classifications hold as GPU vendors, GPU frequency, and other parameters change.
Thus, \DESIGN{} 
helps build more efficient HPC systems.


\begin{acks}
The authors acknowledge the Texas Advanced Computing Center (TACC)~\cite{tacc} at The University of Texas at Austin and Advanced Micro Devices, Inc. under the AMD University Program’s AI \& HPC Cluster.
Both TACC and the AMD HPC Cluster provided computational resources that have contributed to the research results reported within this paper. 
This work was also supported in part by 
NSF grant CNS-2312688, 
an AMD University Program (AUP) grant, and 
the U.S. Department of Energy, Office of Science, Office of Advanced Scientific Computing Research, under Award Number DE-SC-0026036.

\end{acks}

\printbibliography

@INPROCEEDINGS{Aaziz2019proxyqmcpack,
  author={Aaziz, Omar and Cook, Jeanine and Vaughan, Courtenay and Richards, David},
  booktitle={{IEEE International Conference on Cluster Computing}}, 
  series = {CLUSTER},
  title={{Proxy or Imposter? A Method and Case Study to Determine the Answer}}, 
  year={2019},
  volume={},
  number={},
  pages={1-9},
  doi={10.1109/CLUSTER.2019.8891049}
}

@INPROCEEDINGS{Aaziz2018proxyapps,
  author={Aaziz, Omar and Cook, Jeanine and Cook, Jonathan and Juedeman, Tanner and Richards, David and Vaughan, Courtenay},
  booktitle={{IEEE International Conference on Cluster Computing}}, 
  series = {CLUSTER},
  title={{A Methodology for Characterizing the Correspondence Between Real and Proxy Applications}}, 
  year={2018},
  volume={},
  number={},
  pages={190-200},
  doi={10.1109/CLUSTER.2018.00037}
}

@inproceedings{adolf2016fathom,
  title={{Fathom: Reference workloads for modern deep learning methods}},
  author={{Adolf, Robert and Rama, Saketh and Reagen, Brandon and Wei, Gu-Yeon and Brooks, David}},
  booktitle={{IEEE International Symposium on Workload Characterization}},
  series = {IISWC},
  pages={1--10},
  year={2016},
  doi = {10.1109/IISWC.2016.7581275},
  url = {https://doi.ieeecomputersociety.org/10.1109/IISWC.2016.7581275},
  publisher = {IEEE Computer Society},
  address = {Los Alamitos, CA, USA},
  month = {9},
  organization={IEEE}
}

@article{Guerreiro2019dvfs,
  title = {{DVFS-aware application classification to improve GPGPUs energy efficiency}},
  journal = {{Parallel Computing}},
  volume = {83},
  pages = {93-117},
  year = {2019},
  issn = {0167-8191},
  doi = {https://doi.org/10.1016/j.parco.2018.02.001},
  url = {https://www.sciencedirect.com/science/article/pii/S0167819118300243},
  author = {João Guerreiro and Aleksandar Ilic and Nuno Roma and Pedro Tomás},
}

@INPROCEEDINGS{MITSuperCloud,
  author={Tang, Benny J. and Chen, Qiqi and Weiss, Matthew L. and Frey, Nathan C. and McDonald, Joseph and Bestor, David and Yee, Charles and Arcand, William and Bergeron, William and Byun, Chansup and Edelman, Daniel and Houle, Michael and Hubbell, Matthew and Jones, Michael and Kepner, Jeremy and Klein, Anna and Michaleas, Adam and Michaleas, Peter and Milechin, Lauren and Mullen, Julia and Prout, Andrew and Reuther, Albert and Rosa, Antonio and Bowne, Andrew and McEvoy, Lindsey and LI, Baolin and Tiwari, Devesh and Gadepally, Jiay and Samsi, Siddharth},
  booktitle={{IEEE International Parallel and Distributed Processing Symposium Workshops}},
  series = {IPDPSW},
  title={{The MIT Supercloud Workload Classification Challenge}}, 
  year={2022},
  volume={},
  number={},
  pages={708-714},
  doi={10.1109/IPDPSW55747.2022.00122}
}

@INPROCEEDINGS{Antici2024MCBound,
  author={Antici, Francesco and Bartolini, Andrea and Kiziltan, Zeynep and Babaoglu, Ozalp and Kodama, Yuetsu},
  booktitle={{International Conference for High Performance Computing, Networking, Storage and Analysis}},
  series = {SC},
  title={{MCBound: An Online Framework to Characterize and Classify Memory/Compute-bound HPC Jobs}},
  year={2024},
  volume={},
  number={},
  pages={1-15},
  doi={10.1109/SC41406.2024.00062}
}

@INPROCEEDINGS{PMSysScaleSC24,
  author={Karimi, Ahmad Maroof and Maiterth, Matthias and Shin, Woong and Sattar, Naw Safrin and Lu, Hao and Wang, Feiyi},
  booktitle={{Workshops of the International Conference for High Performance Computing, Networking, Storage and Analysis}},
  series = {SC24-W},
  title={{Exploring the Frontiers of Energy Efficiency using Power Management at System Scale}},
  year={2024},
  volume={},
  number={},
  pages={1835-1844},
  doi={10.1109/SCW63240.2024.00230}
}

@ARTICLE{DLRCap2024,
  author={Wang, Yiming and Hao, Meng and He, Hui and Zhang, Weizhe and Tang, Qiuyuan and Sun, Xiaoyang and Wang, Zheng},
  journal={{IEEE Transactions on Sustainable Computing}},
  title={{DRLCAP: Runtime GPU Frequency Capping With Deep Reinforcement Learning}}, 
  year={2024},
  volume={9},
  number={5},
  pages={712-726},
  doi={10.1109/TSUSC.2024.3362697}
}

@inproceedings{jain2024pal,
  title={{PAL: A Variability-Aware Policy for Scheduling ML Workloads in GPU Clusters}},
  author={Jain, Rutwik and Tran, Brandon and Chen, Keting and Sinclair, Matthew D and Venkataraman, Shivaram},
  booktitle={{International Conference for High Performance Computing, Networking, Storage and Analysis}},
  series = {SC},
  pages={1--18},
  year={2024},
  organization={IEEE}
}

@manual{nvidia_nsight_compute,
  title = {{Nsight Compute Documentation}},
  author = {NVIDIA},
  year = {2025},
  note = {Accessed 2025},
  url = {https://docs.nvidia.com/nsight-compute/NsightCompute/index.html}
}

@manual{llnl_coral2_benchmarks,
  title = {{CORAL-2 Benchmarks}},
  author = {Lawrence Livermore National Laboratory},
  year = {2025},
  note = {Accessed 2025},
  url = {https://asc.llnl.gov/coral-2-benchmarks}
}

@manual{mlcommons_benchmarks,
  title = {{MLCommons Benchmarks}},
  author = {MLCommons},
  year = {2025},
  note = {Accessed 2025},
  url = {https://mlcommons.org/benchmarks/}
}

@manual{olcf6_benchmarks,
  title = {{OLCF-6 Technical Requirements: Benchmarks}},
  author = {Oak Ridge Leadership Computing Facility},
  year = {2025},
  note = {Accessed 2025},
  url = {https://www.olcf.ornl.gov/draft-olcf-6-technical-requirements/benchmarks/}
}

@INPROCEEDINGS{FugakuPoints,
  author={Solórzano, Ana Luisa Veroneze and Sato, Kento and Yamamoto, Keiji and Shoji, Fumiyoshi and Brandt, Jim M. and Schwaller, Benjamin and Walton, Sara Petra and Green, Jennifer and Tiwari, Devesh},
  booktitle={{International Conference for High Performance Computing, Networking, Storage and Analysis}}, 
  series = {SC},
  title={{Toward Sustainable HPC: In-Production Deployment of Incentive-Based Power Efficiency Mechanism on the Fugaku Supercomputer}}, 
  year={2024},
  volume={},
  number={},
  pages={1-16},
  doi={10.1109/SC41406.2024.00030}
}

@INPROCEEDINGS{karlin2013lulesh,
  author={Karlin, Ian and Bhatele, Abhinav and Keasler, Jeff and Chamberlain, Bradford L. and Cohen, Jonathan and Devito, Zachary and Haque, Riyaz and Laney, Dan and Luke, Edward and Wang, Felix and Richards, David and Schulz, Martin and Still, Charles H.},
  booktitle={{IEEE 27th International Symposium on Parallel and Distributed Processing}},
  title={{Exploring Traditional and Emerging Parallel Programming Models Using a Proxy Application}}, 
  year={2013},
  volume={},
  number={},
  series = {IPDPS},
  pages={919-932},
  doi={10.1109/IPDPS.2013.115}
}

@misc{touvron2023llama2openfoundation,
  title={{Llama 2: Open Foundation and Fine-Tuned Chat Models}}, 
  author={Hugo Touvron and Louis Martin and Kevin Stone and Peter Albert and Amjad Almahairi and Yasmine Babaei and Nikolay Bashlykov and Soumya Batra and Prajjwal Bhargava and Shruti Bhosale and Dan Bikel and Lukas Blecher and Cristian Canton Ferrer and Moya Chen and Guillem Cucurull and David Esiobu and Jude Fernandes and Jeremy Fu and Wenyin Fu and Brian Fuller and Cynthia Gao and Vedanuj Goswami and Naman Goyal and Anthony Hartshorn and Saghar Hosseini and Rui Hou and Hakan Inan and Marcin Kardas and Viktor Kerkez and Madian Khabsa and Isabel Kloumann and Artem Korenev and Punit Singh Koura and Marie-Anne Lachaux and Thibaut Lavril and Jenya Lee and Diana Liskovich and Yinghai Lu and Yuning Mao and Xavier Martinet and Todor Mihaylov and Pushkar Mishra and Igor Molybog and Yixin Nie and Andrew Poulton and Jeremy Reizenstein and Rashi Rungta and Kalyan Saladi and Alan Schelten and Ruan Silva and Eric Michael Smith and Ranjan Subramanian and Xiaoqing Ellen Tan and Binh Tang and Ross Taylor and Adina Williams and Jian Xiang Kuan and Puxin Xu and Zheng Yan and Iliyan Zarov and Yuchen Zhang and Angela Fan and Melanie Kambadur and Sharan Narang and Aurelien Rodriguez and Robert Stojnic and Sergey Edunov and Thomas Scialom},
  year={2023},
  eprint={2307.09288},
  archivePrefix={arXiv},
  primaryClass={cs.CL},
  url={https://arxiv.org/abs/2307.09288}, 
}

@article{Ahdritz2022openfold,
    author = {Ahdritz, Gustaf and Bouatta, Nazim and Floristean, Christina and Kadyan, Sachin and Xia, Qinghui and Gerecke, William and O{\textquoteright}Donnell, Timothy J and Berenberg, Daniel and Fisk, Ian and Zanichelli, Niccolò and Zhang, Bo and Nowaczynski, Arkadiusz and Wang, Bei and Stepniewska-Dziubinska, Marta M and Zhang, Shang and Ojewole, Adegoke and Guney, Murat Efe and Biderman, Stella and Watkins, Andrew M and Ra, Stephen and Lorenzo, Pablo Ribalta and Nivon, Lucas and Weitzner, Brian and Ban, Yih-En Andrew and Sorger, Peter K and Mostaque, Emad and Zhang, Zhao and Bonneau, Richard and AlQuraishi, Mohammed},
    title = {{{O}pen{F}old: {R}etraining {A}lpha{F}old2 yields new insights into its learning mechanisms and capacity for generalization}},
    elocation-id = {2022.11.20.517210},
    year = {2022},
    doi = {10.1101/2022.11.20.517210},
    publisher = {Cold Spring Harbor Laboratory},
    URL = {https://www.biorxiv.org/content/10.1101/2022.11.20.517210},
    eprint = {https://www.biorxiv.org/content/early/2022/11/22/2022.11.20.517210.full.pdf},
    journal = {bioRxiv}
}

@misc{ahdritz2023openproteinset,
      title={{O}pen{P}rotein{S}et: {T}raining data for structural biology at scale}, 
      author={Gustaf Ahdritz and Nazim Bouatta and Sachin Kadyan and Lukas Jarosch and Daniel Berenberg and Ian Fisk and Andrew M. Watkins and Stephen Ra and Richard Bonneau and Mohammed AlQuraishi},
      year={2023},
      eprint={2308.05326},
      archivePrefix={arXiv},
      primaryClass={q-bio.BM}
}

@article{LAMMPS,
  author  = {A. P. Thompson and H. M. Aktulga and R. Berger and
             D. S. Bolintineanu and W. M. Brown and P. S. Crozier and
             P. J. in 't Veld and A. Kohlmeyer and S. G. Moore and T. D. Nguyen and
             R. Shan and M. J. Stevens and J. Tranchida and C. Trott and S. J. Plimpton},
  title   = {{LAMMPS - A Flexible Simulation tool for Particle-based Materials Modeling at the Atomic, Meso, and Continuum Scales}},
  journal = {Comp. Phys. Comm.},
  volume  = {271},
  pages   = {108171},
  year    = {2022},
  doi     = {10.1016/j.cpc.2021.108171}
}

@manual{olcf6_milc,
  title = {{The MILC Benchmark}},
  author = {{Oak Ridge National Laboratory Leadership Computing Facility}},
  year = {2023},
  note = {Accessed 2025},
  url = {https://www.olcf.ornl.gov/wp-content/uploads/OLCF-6_MILC_description-1.pdf}
}

@manual{olcf6_mpsdns,
  title = {{OLCF-6 Benchmark M-PSDNS Code}},
  author = {{Oak Ridge National Laboratory Leadership Computing Facility}},
  year = {2023},
  note = {Accessed 2025},
  url = {https://www.olcf.ornl.gov/wp-content/uploads/OLCF-6_M-PSDNS_description-1.pdf}
}

@misc{cublas,
  title        = {{cuBLAS}},
  author       = {NVIDIA},
  howpublished = {\url{https://developer.nvidia.com/cublas}},
  year = {2025},
}

@misc{ResNet-pyTorchRef,
  title        = {{ResNet: Deep residual networks pre-trained on ImageNet}},
  howpublished = {\url{https://pytorch.org/hub/pytorch_vision_resnet/}},
  key          = {resnet-pytorch}
}

@article{chen2021rgatrelationalgraphattention,
    title={{r-GAT: Relational Graph Attention Network for Multi-Relational Graphs}}, 
    author={Meiqi Chen and Yuan Zhang and Xiaoyu Kou and Yuntao Li and Yan Zhang},
    year={2021},
    volume = {},
    number = {},
    eprint={2109.05922},
    archivePrefix={arXiv},
    primaryClass={cs.CL},
	journal = {arXiv preprint arXiv:2109.05922},
    numpages = {9},
}

@article{ZengZhang2023-deepmd2,
    author = {{Zeng, Jinzhe and Zhang, Duo and Lu, Denghui and Mo, Pinghui and Li, Zeyu and Chen, Yixiao and Rynik, Marián and Huang, Li’ang and Li, Ziyao and Shi, Shaochen and Wang, Yingze and Ye, Haotian and Tuo, Ping and Yang, Jiabin and Ding, Ye and Li, Yifan and Tisi, Davide and Zeng, Qiyu and Bao, Han and Xia, Yu and Huang, Jiameng and Muraoka, Koki and Wang, Yibo and Chang, Junhan and Yuan, Fengbo and Bore, Sigbjørn Løland and Cai, Chun and Lin, Yinnian and Wang, Bo and Xu, Jiayan and Zhu, Jia-Xin and Luo, Chenxing and Zhang, Yuzhi and Goodall, Rhys E. A. and Liang, Wenshuo and Singh, Anurag Kumar and Yao, Sikai and Zhang, Jingchao and Wentzcovitch, Renata and Han, Jiequn and Liu, Jie and Jia, Weile and York, Darrin M. and E, Weinan and Car, Roberto and Zhang, Linfeng and Wang, Han}},
    title = {{DeePMD-kit v2: A Software Package for Deep Potential Models}},
    journal = {The Journal of Chemical Physics},
    volume = {159},
    number = {5},
    pages = {054801},
    year = {2023},
    month = {08},
    issn = {0021-9606},
    doi = {10.1063/5.0155600},
    url = {https://doi.org/10.1063/5.0155600},
    eprint = {https://pubs.aip.org/aip/jcp/article-pdf/doi/10.1063/5.0155600/18281511/054801\_1\_5.0155600.pdf},
}

@techreport{PageRankSpMV,
  number      = {1999-66},
  author      = {{Lawrence Page and Sergey Brin and Rajeev Motwani and Terry Winograd}},
  note        = {Previous number = SIDL-WP-1999-0120},
  title       = {{The PageRank Citation Ranking: Bringing Order to the Web.}},
  type        = {Technical Report},
  publisher   = {Stanford InfoLab},
  year        = {1999},
  institution = {Stanford InfoLab},
  url         = {http://ilpubs.stanford.edu:8090/422/},
}

@article{PhysRevLett-LSMS,
  title = {{Order-N Multiple Scattering Approach to Electronic Structure Calculations}},
  author = {Wang, Yang and Stocks, G. M. and Shelton, W. A. and Nicholson, D. M. C. and Szotek, Z. and Temmerman, W. M.},
  journal = {Phys. Rev. Lett.},
  volume = {75},
  issue = {15},
  pages = {2867--2870},
  numpages = {0},
  year = {1995},
  publisher = {American Physical Society},
  doi = {10.1103/PhysRevLett.75.2867},
  url = {https://link.aps.org/doi/10.1103/PhysRevLett.75.2867}
}

@article{EISENBACH2017lsms,
  title = {{GPU acceleration of the Locally Selfconsistent Multiple Scattering code for first principles calculation of the ground state and statistical physics of materials}},
  journal = {Computer Physics Communications},
  volume = {211},
  pages = {2-7},
  year = {2017},
  note = {High Performance Computing for Advanced Modeling and Simulation of Materials},
  issn = {0010-4655},
  doi = {10.1016/j.cpc.2016.07.013},
  url = {https://www.sciencedirect.com/science/article/pii/S0010465516301953},
  author = {{Markus Eisenbach and Jeff Larkin and Justin Lutjens and Steven Rennich and James H. Rogers}}
}

@inproceedings{PatelChoukse2024-POLCA,
  author = {Patel, Pratyush and Choukse, Esha and Zhang, Chaojie and Goiri, \'{I}\~{n}igo and Warrier, Brijesh and Mahalingam, Nithish and Bianchini, Ricardo},
  title = {{Characterizing Power Management Opportunities for LLMs in the Cloud}},
  year = {2024},
  isbn = {9798400703867},
  publisher = {Association for Computing Machinery},
  address = {New York, NY, USA},
  url = {https://doi.org/10.1145/3620666.3651329},
  doi = {10.1145/3620666.3651329},
  booktitle = {Proceedings of the 29th ACM International Conference on Architectural Support for Programming Languages and Operating Systems, Volume 3},
  pages = {207–222},
  numpages = {16},
  location = {La Jolla, CA, USA},
  series = {ASPLOS}
}

@inproceedings{Stijkovic25TAPAS,
  author = {Stojkovic, Jovan and Zhang, Chaojie and Goiri, \'{I}\~{n}igo and Choukse, Esha and Qiu, Haoran and Fonseca, Rodrigo and Torrellas, Josep and Bianchini, Ricardo},
  title = {{TAPAS: Thermal- and Power-Aware Scheduling for LLM Inference in Cloud Platforms}},
  year = {2025},
  isbn = {9798400710797},
  publisher = {Association for Computing Machinery},
  address = {New York, NY, USA},
  url = {https://doi.org/10.1145/3676641.3716025},
  doi = {10.1145/3676641.3716025},
  booktitle = {{Proceedings of the 30th ACM International Conference on Architectural Support for Programming Languages and Operating Systems, Volume 2}},
  pages = {1266–1281},
  numpages = {16},
  location = {Rotterdam, Netherlands},
  series = {ASPLOS}
}

@article{lettich2024powerfragmentationawareonlinescheduling,
  title={{Power- and Fragmentation-aware Online Scheduling for GPU Datacenters}}, 
  author={Francesco Lettich and Emanuele Carlini and Franco Maria Nardini and Raffaele Perego and Salvatore Trani},
  year={2024},
  volume = {},
  number = {},
	 eprint={2412.17484},
  archivePrefix={arXiv},
  primaryClass={cs.DC},
  url={https://arxiv.org/abs/2412.17484},
  archivePrefix={arXiv},
  primaryClass={cs.CL},
  journal = {arXiv preprint arXiv:2412.17484},
  numpages = {11},
}

@inproceedings{Sinha-SC22,
  author = {Sinha, Prasoon and Guliani, Akhil and Jain, Rutwik and Tran, Brandon and Sinclair, Matthew D. and Venkataraman, Shivaram},
  title = {{Not All GPUs Are Created Equal: Characterizing Variability in Large-Scale, Accelerator-Rich Systems}},
  year = {2022},
  publisher = {IEEE Press},
  booktitle = {{Proceedings of the International Conference on High Performance Computing, Networking, Storage and Analysis}},
  articleno = {65},
  numpages = {15},
  location = {Dallas, Texas},
  series = {SC}
}

@inproceedings{CoplinBurtscher2016-gpgpuPower,
  author    = {Coplin, Jared and Burtscher, Martin},
  booktitle = {{IEEE International Parallel and Distributed Processing Symposium Workshops}},
  series    = {IPDPSW},
  title     = {{Energy, Power, and Performance Characterization of GPGPU Benchmark Programs}},
  year      = {2016},
  issn      = {},
  pages     = {1190-1199},
  doi       = {10.1109/IPDPSW.2016.164},
  url       = {https://doi.ieeecomputersociety.org/10.1109/IPDPSW.2016.164},
  month     = {05}
}

@inproceedings{Scogland2015-pwrPerspectives,
  author    = {Scogland, Thomas and Azose, Jonathan and Rohr, David and Rivoire, Suzanne and Bates, Natalie and Hackenberg, Daniel},
  title     = {{Node Variability in Large-Scale Power Measurements: Perspectives from the Green500, Top500 and EEHPCWG}},
  year      = {2015},
  isbn      = {9781450337236},
  publisher = {Association for Computing Machinery},
  address   = {New York, NY, USA},
  url       = {https://doi.org/10.1145/2807591.2807653},
  doi       = {10.1145/2807591.2807653},
  booktitle = {Proceedings of the International Conference for High Performance Computing, Networking, Storage and Analysis},
  articleno = {74},
  numpages  = {11},
  location  = {Austin, Texas},
  series    = {SC '15}
}

@misc{nersc10,
  title = {{NERSC-10 Benchmark Suite}},
  year = {2024},
  month = {3},
  howpublished = {\url{https://www.nersc.gov/systems/nersc-10/benchmarks/}},
  author = {{NERSC}},
}

@misc{tacc,
  author       = {TACC},
  title        = {{Texas Advanced Computing Center}},
  howpublished = {\url{https://www.tacc.utexas.edu/}},
  year         = {2024}
}

@inproceedings{CheBeckmann2013-pannotia,
  author    = {Che, Shuai and Beckmann, Bradford M. and Reinhardt, Stephen K. and Skadron, Kevin},
  title     = {{Pannotia: Understanding Irregular GPGPU Graph Applications}},
  year      = {2013},
  booktitle = {{IEEE International Symposium on Workload Characterization}},
  series    = {{IISWC}},
  doi       = {10.1109/IISWC.2013.6704684},
  issn      = {},
  volume    = {},
  number    = {},
  pages     = {185-195},
  month     = {9}
}

@misc{openfold2,
  title = {{OpenFold2: Replicating AlphaFold2 in the Dark}},
  author = {Derevyanko, Georgy and Lamoureux, Guillame and Outeiral, Carlos and Oda, Toshiyuki and Fuchs, Fabian and Mahajan, Sai Pooja and Moult, John and Haas, Juergen and Maragakis, Paul and Ruzmetov, Talant and AlQuraishi, Mohammed},
  year = {2023},
  howpublished = {\url{https://lupoglaz.github.io/OpenFold2/}},
}

@article{Stevens2023-auroraGPT,
  title = {{Argonne's "AuroraGPT" Project}},
  author = {Stevens, Rick}, 
  year = {2023},
  journal = {{Trillion Parameter Consortium Seminar}},
  series = {TPC},
}

@article{fan2021predicting,
  title={{Predicting Orientation-dependent Plastic Susceptibility from Static Structure in Amorphous Solids via Deep Learning}},
  author={Fan, Zhao and Ma, Evan},
  journal={{Nature Communications}},
  volume={12},
  number={1},
  pages={1--13},
  year={2021},
  publisher={Nature Publishing Group}
}

@article{jumper2021highly,
  title={{Highly Accurate Protein Structure Prediction with AlphaFold}},
  author={Jumper, John and Evans, Richard and Pritzel, Alexander and Green, Tim and Figurnov, Michael and Ronneberger, Olaf and Tunyasuvunakool, Kathryn and Bates, Russ and {\v{Z}}{\'\i}dek, Augustin and Potapenko, Anna and Bridgland, Alex and Meyer, Clemens and Kohl, Simon A. A. and Ballard, Andrew J. and Cowie, Andrew and Romera-Paredes, Bernardino and Nikolov, Stanislav and Jain, Rishub and Adler, Jonas and Back, Trevor and Petersen, Stig and Reiman, David and Clancy, Ellen and Zielinski, Michal and Pacholska, Michalina and Berghammer, Tamas and Bodenstein, Sebastian and Silver, David and Vinyals, Oriol and Senior, Andrew W. and Kavukcuoglu, Koray and Kohli, Pushmeet and Hassabis, Demis},
  journal={Nature},
  volume={596},
  number={7873},
  pages={583--589},
  year={2021},
  publisher={Nature Publishing Group}
}

@article{kates2019predicting,
  title={{Predicting Disruptive Instabilities in Controlled Fusion Plasmas through Deep Learning}},
  author={Kates-Harbeck, Julian and Svyatkovskiy, Alexey and Tang, William},
  journal={Nature},
  volume={568},
  number={7753},
  pages={526--531},
  year={2019},
  publisher={Nature Publishing Group}
}

@article{WangZhang2018-deepmd,
  title = {{DeePMD-kit: A Deep Learning Package for Many-body Potential Energy Representation and Molecular Dynamics}},
  journal = {{Computer Physics Communications}},
  volume = {228},
  pages = {178-184},
  year = {2018},
  issn = {0010-4655},
  doi = {10.1016/j.cpc.2018.03.016},
  url = {https://www.sciencedirect.com/science/article/pii/S0010465518300882},
  author = {Wang, Han and Zhang, Linfeng and Han, Jiequn and E, Weinan}
}

@misc{kecklerpicojoule,
  author={Keckler, Stephen W.},
  title={{Life After Dennard and How I Learned to Love the Picojoule}},
  howpublished={Keynote at MICRO},
  year = {2011},
}

@inproceedings{SmithLoh2024-mi300A,
  author       = {Alan Smith and
                  Gabriel H. Loh and
                  Michael J. Schulte and
                  Mike Ignatowski and
                  Samuel Naffziger and
                  Mike Mantor and
                  Nathan Kalyanasundharam and
                  Vamsi Alla and
                  Nicholas Malaya and
                  Joseph L. Greathouse and
                  Eric Chapman and
                  Raja Swaminathan},
  title        = {{Realizing the AMD Exascale Heterogeneous Processor Vision : Industry
                  Product}},
  booktitle    = {{51st ACM/IEEE Annual International Symposium on Computer Architecture}},
  series       = {ISCA},
  pages        = {876--889},
  publisher    = {{IEEE}},
  address = {Piscataway, NJ, USA},
  year         = {2024},
  url          = {https://doi.org/10.1109/ISCA59077.2024.00068},
  doi          = {10.1109/ISCA59077.2024.00068},
  bibsource    = {dblp computer science bibliography, https://dblp.org}
}

@inproceedings{LohSchulte2023-mi250,
  author = {Loh, Gabriel H. and Schulte, Michael J. and Ignatowski, Mike and Adhinarayanan, Vignesh and Aga, Shaizeen and Aguren, Derrick and Agrawal, Varun and Aji, Ashwin M. and Alsop, Johnathan and Bauman, Paul and Beckmann, Bradford M. and Beigi, Majed Valad and Blagodurov, Sergey and Boraten, Travis and Boyer, Michael and Brantley, William C. and Chalmers, Noel and Chen, Shaoming and Cheng, Kevin and Chu, Michael L. and Cownie, David and Curtis, Nicholas and Del Pino, Joris and Duong, Nam and Duundefinedu, Alexandru and Eckert, Yasuko and Erb, Christopher and Freitag, Chip and Greathouse, Joseph L. and Gurumurthi, Sudhanva and Gutierrez, Anthony and Hamidouche, Khaled and Hossamani, Sachin and Huang, Wei and Islam, Mahzabeen and Jayasena, Nuwan and Kalamatianos, John and Kayiran, Onur and Kotra, Jagadish and Lee, Alan and Lowell, Daniel and Madan, Niti and Majumdar, Abhinandan and Malaya, Nicholas and Manne, Srilatha and Mashimo, Susumu and McDougall, Damon and Mednick, Elliot and Mishkin, Michael and Nutter, Mark and Paul, Indrani and Poremba, Matthew and Potter, Brandon and Punniyamurthy, Kishore and Puthoor, Sooraj and Raasch, Steven E. and Rao, Karthik and Rodgers, Gregory and Scrbak, Marko and Seyedzadeh, Mohammad and Slice, John and Sridharan, Vilas and van Oostrum, Ren\'{e} and van Tassell, Eric and Vishnu, Abhinav and Wasmundt, Samuel and Wilkening, Mark and Wolfe, Noah and Wyse, Mark and Yalavarti, Adithya and Yudanov, Dmitri},
  title = {{A Research Retrospective on AMD's Exascale Computing Journey}},
  year = {2023},
  isbn = {9798400700958},
  publisher = {Association for Computing Machinery},
  address = {New York, NY, USA},
  url = {https://doi.org/10.1145/3579371.3589349},
  doi = {10.1145/3579371.3589349},
  booktitle = {{Proceedings of the 50th Annual International Symposium on Computer Architecture}},
  articleno = {81},
  numpages = {14},
  location = {Orlando, FL, USA},
  series = {ISCA}
}

@INPROCEEDINGS{Gene-GreathouseHPCA15,
  author={{Wu, Gene and Greathouse, Joseph L. and Lyashevsky, Alexander and Jayasena, Nuwan and Chiou, Derek}},
  booktitle={{IEEE 21st International Symposium on High Performance Computer Architecture}},
  series = {HPCA},
  title={GPGPU performance and power estimation using machine learning}, 
  year={2015},
  volume={},
  number={},
  pages={564-576},
  doi={10.1109/HPCA.2015.7056063}
}

@inproceedings{SimplePower-Vijaykrishnan-ISCA00,
  author = {Vijaykrishnan, N. and Kandemir, M. and Irwin, M. J. and Kim, H. S. and Ye, W.},
  title = {{Energy-driven Integrated Hardware-software Optimizations Using SimplePower}},
  year = {2000},
  isbn = {1581132328},
  publisher = {Association for Computing Machinery},
  address = {New York, NY, USA},
  url = {https://doi.org/10.1145/339647.339659},
  doi = {10.1145/339647.339659},
  booktitle = {Proceedings of the 27th Annual International Symposium on Computer Architecture},
  pages = {95–106},
  numpages = {12},
  location = {Vancouver, British Columbia, Canada},
  series = {ISCA '00}
}

@ARTICLE{Choquette2023-hopper,
  author={Choquette, Jack},
  journal={IEEE Micro},
  title={{NVIDIA Hopper H100 GPU: Scaling Performance}},
  year={2023},
  volume={43},
  number={03},
  ISSN={1937-4143},
  pages={9-17},
  doi={10.1109/MM.2023.3256796},
  url = {https://doi.ieeecomputersociety.org/10.1109/MM.2023.3256796},
  publisher={IEEE Computer Society},
  address={Los Alamitos, CA, USA},
  month=may
}

@misc{blackwell,
  title = {{NVIDIA Blackwell Architecture Technical Brief}},
  author = {{NVIDIA}},
  howpublished = {\url{https://resources.nvidia.com/en-us-blackwell-architecture/blackwell-architecture-technical-brief}},
  year = {2024},
}

@ARTICLE{ChoquetteGiroux2018-volta,
  author={Choquette, Jack and Giroux, Olivier and Foley, Denis},
  journal={IEEE Micro},
  title={{Volta: Performance and Programmability}},
  year={2018},
  volume={38},
  number={2},
  pages={42-52},
  doi={10.1109/MM.2018.022071134}
}

@misc{ampere,
  title = {{NVIDIA Ampere Architecture In-Depth}},
  author = {Krashinsky, Ronny and Giroux, Olivier and Jones, Stephen and Stam, Nick and Ramaswamy, Sridhar},
  year = {2020},
  howpublished = {\url{https://developer.nvidia.com/blog/nvidia-ampere-architecture-in-depth/}},
}

@inproceedings{Huang2025-gtcKeynote,
  author = {Huang, Jen-sen},
  title = {{GTC March 2025 Keynote}},
  year = {2025},
  booktitle="{Proceedings of GPU Technology Conference}",
  series = {GTC},
}

@misc{rsmiapi,
  title        = {{ROCm System Management Interface (ROCm SMI) library}},
  author       = {{Advanced Micro Devices, Inc.}},
  year         = {2025},
  howpublished = {\url{https://rocmdocs.amd.com/projects/rocm_smi_lib/en/latest/index.html}}
}

@techreport{CarterFeddema2023-aiScience,
  author       = {Carter, Jonathan and Feddema, John and Kothe, Doug and Neely, Rob and Pruet, Jason and Stevens, Rick and Balaprakash, Prasanna and Beckman, Pete and Foster, Ian and Iskra, Kamil and Ramanathan, Arvind and Taylor, Valerie and Thakur, Rajeev and Agarwal, Deb and Crivelli, Silvia and de Jong, Bert and Rouson, Damian and Sohn, Mike and Wetter, Michael and Wild, Stefan and Bremer, Timo and Goldman, Michael and Kupresanin, Ana and Peterson, Luc and Spears, Brian and Stevens, Dave and Van Essen, Brian and Bent, Russell and Grosskopf, Mike and Lawrence, Earl and Shipman, Galen and Rose, Kelly and Grout, Ray and Kouakpaizan, Nicholson and Omitaomu, Femi and Peles, Slaven and Ramuhalli, Pradeep and Shankar, Arjun and Womble, David and Zhang, Guannan and Catanach, Tommie and Oldfield, Ron and Rajamanickam, Sivasankaran and Ray, Jaideep and Leung, Mary Ann and Catlett, Charles and Dietrich, Emily M.},
  title        = {{Advanced Research Directions on AI for Science, Energy, and Security: Report on Summer 2022 Workshops}},
  institution  = {Argonne National Laboratory (ANL), Argonne, IL (United States)},
  doi          = {10.2172/1986455},
  url          = {https://www.osti.gov/biblio/1986455},
  place        = {United States},
  year         = {2023},
  month        = {05}
}

@misc{GanRanganathan2025-googlePowerSwings,
  title = {{Balance of Power: A Full-stack Approach to Power and Thermal Fluctuations in ML Infrastructure}},
  author = {Gan, Houle and Ranganathan, Parthasarathy},
  year = {2025},
  month = {02},
  howpublished = {\url{https://cloud.google.com/blog/topics/systems/mitigating-power-and-thermal-fluctuations-in-ml-infrastructure}},
}

@inproceedings{PigaNarayanan2024-metaDVFS,
  author = {Piga, Leonardo and Narayanan, Iyswarya and Sundarrajan, Aditya and Skach, Matt and Deng, Qingyuan and Maity, Biswadip and Chakkaravarthy, Manoj and Huang, Alison and Dhanotia, Abhishek and Malani, Parth},
  title = {{Expanding Datacenter Capacity with DVFS Boosting: A Safe and Scalable Deployment Experience}},
  year = {2024},
  isbn = {9798400703720},
  publisher = {Association for Computing Machinery},
  address = {New York, NY, USA},
  url = {https://doi.org/10.1145/3617232.3624853},
  doi = {10.1145/3617232.3624853},
  booktitle = {{Proceedings of the 29th ACM International Conference on Architectural Support for Programming Languages and Operating Systems, Volume 1}},
  pages = {150–165},
  numpages = {16},
  location = {La Jolla, CA, USA},
  series = {ASPLOS}
}

@misc{ocp-spec,
  title={{OCP Universal Baseboard (UBB) Design Specification v1.5}},
  author={{OCP Open Accelerator Infrastructure (OAI) Workstreams}},
  year={2022},
  note="\url{https://www.opencompute.org/documents/universal-baseboard-design-specification-v1p5-final-20220223-docx-pdf}",    
}

@misc{mi60,
  title        = {{AMD Radeon Instinct MI60}},
  author       = {AMD},
  howpublished = {\url{https://www.amd.com/system/files/documents/radeon-instinct-mi60-datasheet.pdf}},
  year         = {2018}
}

@inproceedings{BharadwajDas2024-gpuDVFSPredict,
  author = {Bharadwaj, Srikant and Das, Shomit and Mazumdar, Kaushik and Beckmann, Bradford M. and Kosonocky, Stephen},
  title = {{Predict; Don't React for Enabling Efficient Fine-Grain DVFS in GPUs}},
  year = {2024},
  isbn = {9798400703942},
  publisher = {Association for Computing Machinery},
  address = {New York, NY, USA},
  url = {https://doi.org/10.1145/3623278.3624756},
  doi = {10.1145/3623278.3624756},
  booktitle = {{Proceedings of the 28th ACM International Conference on Architectural Support for Programming Languages and Operating Systems, Volume 4}},
  pages = {253–267},
  numpages = {15},
  location = {Vancouver, BC, Canada},
  series = {ASPLOS '23}
}

@INPROCEEDINGS{GeVogt2013-dvfsKepler,
  author={R. {Ge} and R. {Vogt} and J. {Majumder} and A. {Alam} and M. {Burtscher} and Z. {Zong}},
  title={{Effects of Dynamic Voltage and Frequency Scaling on a K20 GPU}},
  year={2013},
  volume={},
  number={},
  pages={826-833},
  doi={10.1109/ICPP.2013.98},
  booktitle={{42nd International Conference on Parallel Processing}},
  series = {ICPP},
}

@INPROCEEDINGS{MeinerzhagenTokunaga2018-gpuDVFS,
  author={Meinerzhagen, Pascal and Tokunaga, Carlos and Malavasi, Andres and Vaidya, Vaibhav and Mendon, Ashwin and Mathaikutty, Deepak and Kulkarni, Jaydeep and Augustine, Charles and Cho, Minki and Ki\
m, Stephen and Matthew, George and Jain, Rinkle and Ryan, Joseph and Peng, Chung-Ching and Paul, Somnath and Vangal, Sriram and Esparza, Brando Perez and Cuellar, Luis and Woodman, Michael and Iyer, Bal\
a and Maiyuran, Subramaniam and Chinya, Gautham and Zou, Chris and Liao, Yuyun and Ravichandran, Krishnan and Wang, Hong and Khellah, Muhammad and Tschanz, James and De, Vivek},
  booktitle={{IEEE International Solid - State Circuits Conference}},
  series = {ISSCC},
  title={{An energy-efficient graphics processor featuring fine-grain DVFS with integrated voltage regulators, execution-unit turbo, and retentive sleep in 14nm tri-gate CMOS}},
  year={2018},
  volume={},
  number={},
  pages={38-40},
  doi={10.1109/ISSCC.2018.8310172}
}

@inproceedings{NathTullsen2015-crisp,
  author = {Nath, Rajib and Tullsen, Dean},
  title = {{The CRISP Performance Model for Dynamic Voltage and Frequency Scaling in a GPGPU}},
  year = {2015},
  isbn = {9781450340342},
  publisher = {Association for Computing Machinery},
  address = {New York, NY, USA},
  url = {https://doi.org/10.1145/2830772.2830826},
  doi = {10.1145/2830772.2830826},
  booktitle = {{Proceedings of the 48th International Symposium on Microarchitecture}},
  pages = {281–293},
  numpages = {13},
  location = {Waikiki, Hawaii},
  series = {MICRO}
}

@inproceedings{MeiYung2013-gpuDVFSMeasure,
  author = {Mei, Xinxin and Yung, Ling Sing and Zhao, Kaiyong and Chu, Xiaowen},
  title = {{A Measurement Study of GPU DVFS on Energy Conservation}},
  year = {2013},
  isbn = {9781450324588},
  publisher = {Association for Computing Machinery},
  address = {New York, NY, USA},
  url = {https://doi.org/10.1145/2525526.2525852},
  doi = {10.1145/2525526.2525852},
  booktitle = {{Proceedings of the Workshop on Power-Aware Computing and Systems}},
  articleno = {10},
  numpages = {5},
  location = {Farmington, Pennsylvania},
  series = {HotPower}
}

@inproceedings{AvalosKhairy2021-pka,
  author = {Avalos Baddouh, Cesar and Khairy, Mahmoud and Green, Roland N. and Payer, Mathias and Rogers, Timothy G.},
  title = {{Principal Kernel Analysis: A Tractable Methodology to Simulate Scaled GPU Workloads}},
  year = {2021},
  isbn = {9781450385572},
  publisher = {Association for Computing Machinery},
  address = {New York, NY, USA},
  url = {https://doi.org/10.1145/3466752.3480100},
  doi = {10.1145/3466752.3480100},
  booktitle = {{54th Annual IEEE/ACM International Symposium on Microarchitecture}},
  pages = {724–737},
  numpages = {14},
  location = {Virtual Event, Greece},
  series = {MICRO}
}

@inproceedings{PatiAga20-seqPoints,
  author = {Pati, Suchita and Aga, Shaizeen and Sinclair, Matthew D. and Jayasena, Nuwan},
  title = {{SeqPoint: Identifying Representative Iterations of Sequence-based Neural Networks}},
  booktitle = {{IEEE International Symposium on Performance Analysis of Systems and Software}},
  series = {ISPASS},
  year = {2020},
  month = {8},
  publisher = {IEEE Computer Society},
  address = {Washington, DC, USA},
  volume={},
  number={},
  pages={69-80},
  doi={10.1109/ISPASS48437.2020.00017}
}

@INPROCEEDINGS{WunderlichWenisch2003-smarts,
  author={Wunderlich, R.E. and Wenisch, T.F. and Falsafi, B. and Hoe, J.C.},
  booktitle={{Proceedings of 30th Annual International Symposium on Computer Architecture}}, 
  series = {ISCA},
  title={{SMARTS: Accelerating Microarchitecture Simulation via Rigorous Statistical Sampling}}, 
  year={2003},
  volume={},
  number={},
  pages={84-95},
  doi={10.1109/ISCA.2003.1206991}
}

@inproceedings{Sherwood02,
 author = {Sherwood, Timothy and Perelman, Erez and Hamerly, Greg and Calder, Brad},
 title = {{Automatically Characterizing Large Scale Program Behavior}},
 booktitle = {{Proceedings of the 10th International Conference on Architectural Support for Programming Languages and Operating Systems}},
 series = {ASPLOS},
 year = {2002},
}

@inproceedings{Sherwood01,
  title={{Basic Block Distribution Analysis to Find Periodic Behavior and Simulation Points in Applications}},
  author={Sherwood, Timothy and Perelman, Erez and Calder, Brad},
  booktitle={{Proceedings of the 2001 International Conference on Parallel Architectures and Compilation Techniques}},
  pages={3--14},
  year={2001},
  series = {PACT},
  organization={IEEE}
}

@INPROCEEDINGS {Patel2024Splitwise,
author = { Patel, Pratyush and Choukse, Esha and Zhang, Chaojie and Shah, Aashaka and Goiri, Inigo and Maleki, Saeed and Bianchini, Ricardo },
booktitle = {{ACM/IEEE 51st Annual International Symposium on Computer Architecture}},
series = {ISCA},
title = {{Splitwise: Efficient Generative LLM Inference Using Phase Splitting }},
year = {2024},
volume = {},
ISSN = {},
pages = {118-132},
doi = {10.1109/ISCA59077.2024.00019},
url = {https://doi.ieeecomputersociety.org/10.1109/ISCA59077.2024.00019},
publisher = {IEEE Computer Society},
address = {Los Alamitos, CA, USA},
month =Jul}

@article{XIA201539,
title = {{Learning Similarity with Cosine Similarity Ensemble}},
journal = {Information Sciences},
volume = {307},
pages = {39-52},
year = {2015},
issn = {0020-0255},
doi = {https://doi.org/10.1016/j.ins.2015.02.024},
url = {https://www.sciencedirect.com/science/article/pii/S0020025515001243},
author = {Peipei Xia and Li Zhang and Fanzhang Li},
}

@misc{cosinedistance,
  author = {{Scikit Learn}},
  title={{API Reference: sklearn.metrics: cosine distances}},
  url={https://scikit-learn.org/stable/modules/generated/sklearn.metrics.pairwise.cosine_distances.html},
  year={2007},      
}

@INPROCEEDINGS{YangNvidiaPowerSensor,
  author={Yang, Zeyu and Adamek, Karel and Armour, Wesley},
  booktitle={{International Conference for High Performance Computing, Networking, Storage and Analysis}},
  series = {SC},
  title={{Accurate and Convenient Energy Measurements for GPUs: A Detailed Study of NVIDIA GPU’s Built-In Power Sensor}}, 
  year={2024},
  volume={},
  number={},
  pages={1-17},
  doi={10.1109/SC41406.2024.00028}
}

@article{Wang2017Gunrock,
  author = {Wang, Yangzihao and Pan, Yuechao and Davidson, Andrew and Wu, Yuduo and Yang, Carl and Wang, Leyuan and Osama, Muhammad and Yuan, Chenshan and Liu, Weitang and Riffel, Andy T. and Owens, John D.},
  title = {{Gunrock: GPU Graph Analytics}},
  year = {2017},
  issue_date = {March 2017},
  publisher = {Association for Computing Machinery},
  address = {New York, NY, USA},
  volume = {4},
  number = {1},
  issn = {2329-4949},
  url = {https://doi.org/10.1145/3108140},
  doi = {10.1145/3108140},
  journal = {ACM Trans. Parallel Comput.},
  month = aug,
  articleno = {3},
  numpages = {49}
}

@misc{grattafiori2024llama3herdmodels,
      title={{The Llama 3 Herd of Models}}, 
      author={Aaron Grattafiori and Abhimanyu Dubey and Abhinav Jauhri and Abhinav Pandey and Abhishek Kadian and Ahmad Al-Dahle and Aiesha Letman and Akhil Mathur and Alan Schelten and Alex Vaughan and Amy Yang and Angela Fan and Anirudh Goyal and Anthony Hartshorn and Aobo Yang and Archi Mitra and Archie Sravankumar and Artem Korenev and Arthur Hinsvark and Arun Rao and Aston Zhang and Aurelien Rodriguez and Austen Gregerson and Ava Spataru and Baptiste Roziere and Bethany Biron and Binh Tang and Bobbie Chern and Charlotte Caucheteux and Chaya Nayak and Chloe Bi and Chris Marra and Chris McConnell and Christian Keller and Christophe Touret and Chunyang Wu and Corinne Wong and Cristian Canton Ferrer and Cyrus Nikolaidis and Damien Allonsius and Daniel Song and Danielle Pintz and Danny Livshits and Danny Wyatt and David Esiobu and Dhruv Choudhary and Dhruv Mahajan and Diego Garcia-Olano and Diego Perino and Dieuwke Hupkes and Egor Lakomkin and Ehab AlBadawy and Elina Lobanova and Emily Dinan and Eric Michael Smith and Filip Radenovic and Francisco Guzmán and Frank Zhang and Gabriel Synnaeve and Gabrielle Lee and Georgia Lewis Anderson and Govind Thattai and Graeme Nail and Gregoire Mialon and Guan Pang and Guillem Cucurell and Hailey Nguyen and Hannah Korevaar and Hu Xu and Hugo Touvron and Iliyan Zarov and Imanol Arrieta Ibarra and Isabel Kloumann and Ishan Misra and Ivan Evtimov and Jack Zhang and Jade Copet and Jaewon Lee and Jan Geffert and Jana Vranes and Jason Park and Jay Mahadeokar and Jeet Shah and Jelmer van der Linde and Jennifer Billock and Jenny Hong and Jenya Lee and Jeremy Fu and Jianfeng Chi and Jianyu Huang and Jiawen Liu and Jie Wang and Jiecao Yu and Joanna Bitton and Joe Spisak and Jongsoo Park and Joseph Rocca and Joshua Johnstun and Joshua Saxe and Junteng Jia and Kalyan Vasuden Alwala and Karthik Prasad and Kartikeya Upasani and Kate Plawiak and Ke Li and Kenneth Heafield and Kevin Stone and Khalid El-Arini and Krithika Iyer and Kshitiz Malik and Kuenley Chiu and Kunal Bhalla and Kushal Lakhotia and Lauren Rantala-Yeary and Laurens van der Maaten and Lawrence Chen and Liang Tan and Liz Jenkins and Louis Martin and Lovish Madaan and Lubo Malo and Lukas Blecher and Lukas Landzaat and Luke de Oliveira and Madeline Muzzi and Mahesh Pasupuleti and Mannat Singh and Manohar Paluri and Marcin Kardas and Maria Tsimpoukelli and Mathew Oldham and Mathieu Rita and Maya Pavlova and Melanie Kambadur and Mike Lewis and Min Si and Mitesh Kumar Singh and Mona Hassan and Naman Goyal and Narjes Torabi and Nikolay Bashlykov and Nikolay Bogoychev and Niladri Chatterji and Ning Zhang and Olivier Duchenne and Onur Çelebi and Patrick Alrassy and Pengchuan Zhang and Pengwei Li and Petar Vasic and Peter Weng and Prajjwal Bhargava and Pratik Dubal and Praveen Krishnan and Punit Singh Koura and Puxin Xu and Qing He and Qingxiao Dong and Ragavan Srinivasan and Raj Ganapathy and Ramon Calderer and Ricardo Silveira Cabral and Robert Stojnic and Roberta Raileanu and Rohan Maheswari and Rohit Girdhar and Rohit Patel and Romain Sauvestre and Ronnie Polidoro and Roshan Sumbaly and Ross Taylor and Ruan Silva and Rui Hou and Rui Wang and Saghar Hosseini and Sahana Chennabasappa and Sanjay Singh and Sean Bell and Seohyun Sonia Kim and Sergey Edunov and Shaoliang Nie and Sharan Narang and Sharath Raparthy and Sheng Shen and Shengye Wan and Shruti Bhosale and Shun Zhang and Simon Vandenhende and Soumya Batra and Spencer Whitman and Sten Sootla and Stephane Collot and Suchin Gururangan and Sydney Borodinsky and Tamar Herman and Tara Fowler and Tarek Sheasha and Thomas Georgiou and Thomas Scialom and Tobias Speckbacher and Todor Mihaylov and Tong Xiao and Ujjwal Karn and Vedanuj Goswami and Vibhor Gupta and Vignesh Ramanathan and Viktor Kerkez and Vincent Gonguet and Virginie Do and Vish Vogeti and Vítor Albiero and Vladan Petrovic and Weiwei Chu and Wenhan Xiong and Wenyin Fu and Whitney Meers and Xavier Martinet and Xiaodong Wang and Xiaofang Wang and Xiaoqing Ellen Tan and Xide Xia and Xinfeng Xie and Xuchao Jia and Xuewei Wang and Yaelle Goldschlag and Yashesh Gaur and Yasmine Babaei and Yi Wen and Yiwen Song and Yuchen Zhang and Yue Li and Yuning Mao and Zacharie Delpierre Coudert and Zheng Yan and Zhengxing Chen and Zoe Papakipos and Aaditya Singh and Aayushi Srivastava and Abha Jain and Adam Kelsey and Adam Shajnfeld and Adithya Gangidi and Adolfo Victoria and Ahuva Goldstand and Ajay Menon and Ajay Sharma and Alex Boesenberg and Alexei Baevski and Allie Feinstein and Amanda Kallet and Amit Sangani and Amos Teo and Anam Yunus and Andrei Lupu and Andres Alvarado and Andrew Caples and Andrew Gu and Andrew Ho and Andrew Poulton and Andrew Ryan and Ankit Ramchandani and Annie Dong and Annie Franco and Anuj Goyal and Aparajita Saraf and Arkabandhu Chowdhury and Ashley Gabriel and Ashwin Bharambe and Assaf Eisenman and Azadeh Yazdan and Beau James and Ben Maurer and Benjamin Leonhardi and Bernie Huang and Beth Loyd and Beto De Paola and Bhargavi Paranjape and Bing Liu and Bo Wu and Boyu Ni and Braden Hancock and Bram Wasti and Brandon Spence and Brani Stojkovic and Brian Gamido and Britt Montalvo and Carl Parker and Carly Burton and Catalina Mejia and Ce Liu and Changhan Wang and Changkyu Kim and Chao Zhou and Chester Hu and Ching-Hsiang Chu and Chris Cai and Chris Tindal and Christoph Feichtenhofer and Cynthia Gao and Damon Civin and Dana Beaty and Daniel Kreymer and Daniel Li and David Adkins and David Xu and Davide Testuggine and Delia David and Devi Parikh and Diana Liskovich and Didem Foss and Dingkang Wang and Duc Le and Dustin Holland and Edward Dowling and Eissa Jamil and Elaine Montgomery and Eleonora Presani and Emily Hahn and Emily Wood and Eric-Tuan Le and Erik Brinkman and Esteban Arcaute and Evan Dunbar and Evan Smothers and Fei Sun and Felix Kreuk and Feng Tian and Filippos Kokkinos and Firat Ozgenel and Francesco Caggioni and Frank Kanayet and Frank Seide and Gabriela Medina Florez and Gabriella Schwarz and Gada Badeer and Georgia Swee and Gil Halpern and Grant Herman and Grigory Sizov and Guangyi and Zhang and Guna Lakshminarayanan and Hakan Inan and Hamid Shojanazeri and Han Zou and Hannah Wang and Hanwen Zha and Haroun Habeeb and Harrison Rudolph and Helen Suk and Henry Aspegren and Hunter Goldman and Hongyuan Zhan and Ibrahim Damlaj and Igor Molybog and Igor Tufanov and Ilias Leontiadis and Irina-Elena Veliche and Itai Gat and Jake Weissman and James Geboski and James Kohli and Janice Lam and Japhet Asher and Jean-Baptiste Gaya and Jeff Marcus and Jeff Tang and Jennifer Chan and Jenny Zhen and Jeremy Reizenstein and Jeremy Teboul and Jessica Zhong and Jian Jin and Jingyi Yang and Joe Cummings and Jon Carvill and Jon Shepard and Jonathan McPhie and Jonathan Torres and Josh Ginsburg and Junjie Wang and Kai Wu and Kam Hou U and Karan Saxena and Kartikay Khandelwal and Katayoun Zand and Kathy Matosich and Kaushik Veeraraghavan and Kelly Michelena and Keqian Li and Kiran Jagadeesh and Kun Huang and Kunal Chawla and Kyle Huang and Lailin Chen and Lakshya Garg and Lavender A and Leandro Silva and Lee Bell and Lei Zhang and Liangpeng Guo and Licheng Yu and Liron Moshkovich and Luca Wehrstedt and Madian Khabsa and Manav Avalani and Manish Bhatt and Martynas Mankus and Matan Hasson and Matthew Lennie and Matthias Reso and Maxim Groshev and Maxim Naumov and Maya Lathi and Meghan Keneally and Miao Liu and Michael L. Seltzer and Michal Valko and Michelle Restrepo and Mihir Patel and Mik Vyatskov and Mikayel Samvelyan and Mike Clark and Mike Macey and Mike Wang and Miquel Jubert Hermoso and Mo Metanat and Mohammad Rastegari and Munish Bansal and Nandhini Santhanam and Natascha Parks and Natasha White and Navyata Bawa and Nayan Singhal and Nick Egebo and Nicolas Usunier and Nikhil Mehta and Nikolay Pavlovich Laptev and Ning Dong and Norman Cheng and Oleg Chernoguz and Olivia Hart and Omkar Salpekar and Ozlem Kalinli and Parkin Kent and Parth Parekh and Paul Saab and Pavan Balaji and Pedro Rittner and Philip Bontrager and Pierre Roux and Piotr Dollar and Polina Zvyagina and Prashant Ratanchandani and Pritish Yuvraj and Qian Liang and Rachad Alao and Rachel Rodriguez and Rafi Ayub and Raghotham Murthy and Raghu Nayani and Rahul Mitra and Rangaprabhu Parthasarathy and Raymond Li and Rebekkah Hogan and Robin Battey and Rocky Wang and Russ Howes and Ruty Rinott and Sachin Mehta and Sachin Siby and Sai Jayesh Bondu and Samyak Datta and Sara Chugh and Sara Hunt and Sargun Dhillon and Sasha Sidorov and Satadru Pan and Saurabh Mahajan and Saurabh Verma and Seiji Yamamoto and Sharadh Ramaswamy and Shaun Lindsay and Shaun Lindsay and Sheng Feng and Shenghao Lin and Shengxin Cindy Zha and Shishir Patil and Shiva Shankar and Shuqiang Zhang and Shuqiang Zhang and Sinong Wang and Sneha Agarwal and Soji Sajuyigbe and Soumith Chintala and Stephanie Max and Stephen Chen and Steve Kehoe and Steve Satterfield and Sudarshan Govindaprasad and Sumit Gupta and Summer Deng and Sungmin Cho and Sunny Virk and Suraj Subramanian and Sy Choudhury and Sydney Goldman and Tal Remez and Tamar Glaser and Tamara Best and Thilo Koehler and Thomas Robinson and Tianhe Li and Tianjun Zhang and Tim Matthews and Timothy Chou and Tzook Shaked and Varun Vontimitta and Victoria Ajayi and Victoria Montanez and Vijai Mohan and Vinay Satish Kumar and Vishal Mangla and Vlad Ionescu and Vlad Poenaru and Vlad Tiberiu Mihailescu and Vladimir Ivanov and Wei Li and Wenchen Wang and Wenwen Jiang and Wes Bouaziz and Will Constable and Xiaocheng Tang and Xiaojian Wu and Xiaolan Wang and Xilun Wu and Xinbo Gao and Yaniv Kleinman and Yanjun Chen and Ye Hu and Ye Jia and Ye Qi and Yenda Li and Yilin Zhang and Ying Zhang and Yossi Adi and Youngjin Nam and Yu and Wang and Yu Zhao and Yuchen Hao and Yundi Qian and Yunlu Li and Yuzi He and Zach Rait and Zachary DeVito and Zef Rosnbrick and Zhaoduo Wen and Zhenyu Yang and Zhiwei Zhao and Zhiyu Ma},
      year={2024},
      eprint={2407.21783},
      archivePrefix={arXiv},
      primaryClass={cs.AI},
      url={https://arxiv.org/abs/2407.21783}, 
}

@misc{podell2023sdxlimprovinglatentdiffusion,
      title={{SDXL: Improving Latent Diffusion Models for High-Resolution Image Synthesis}}, 
      author={Dustin Podell and Zion English and Kyle Lacey and Andreas Blattmann and Tim Dockhorn and Jonas Müller and Joe Penna and Robin Rombach},
      year={2023},
      eprint={2307.01952},
      archivePrefix={arXiv},
      primaryClass={cs.CV},
      url={https://arxiv.org/abs/2307.01952}, 
}

@misc{mullner2011modernhierarchicalagglomerativeclustering,
      title={{Modern Hierarchical, Agglomerative Clustering Algorithms}}, 
      author={{Daniel Müllner}},
      year={2011},
      eprint={1109.2378},
      archivePrefix={arXiv},
      primaryClass={stat.ML},
      url={https://arxiv.org/abs/1109.2378}, 
}

@inproceedings{DingZheng2023-mirage,
  author = {Ding, Qiyang and Zheng, Pengfei and Kudari, Shreyas and Venkataraman, Shivaram and Zhang, Zhao},
  title = {{Mirage: Towards Low-interruption Services on Batch GPU Clusters with Reinforcement Learning}},
  year = {2023},
  isbn = {9798400701092},
  publisher = {Association for Computing Machinery},
  address = {New York, NY, USA},
  url = {https://doi.org/10.1145/3581784.3607042},
  doi = {10.1145/3581784.3607042},
  booktitle = {{Proceedings of the International Conference for High Performance Computing, Networking, Storage and Analysis}},
  articleno = {25},
  numpages = {13},
  location = {Denver, CO, USA},
  series = {SC}
}

@article{GholamiYao2024-MLGrowth,
  author={Gholami, Amir and Yao, Zhewei and Kim, Sehoon and Hooper, Coleman and Mahoney, Michael W. and Keutzer, Kurt},
  journal={IEEE Micro},
  title={{AI and Memory Wall}},
  year={2024},
  volume={44},
  number={03},
}

@techreport{ShehabiNewkirk2024-doeDCEnergy,
  author={Shehabi, Arman and Newkirk, Alex and Smith, Sarah J and Hubbard, Alex and Lei, Nuoa and Siddik, Md Abu Bakar and Holecek, Billie and Koomey, Jonathan and Masanet, Eric and Sartor, Dale},
  title = {{2024 United States Data Center Energy Usage Report}},
  institution = {{Lawrence Berkeley National Laboratory}},
  year = {2024},
  month = {12},
  number = {LBNL-2001637},
  location = {Berkeley, CA},
}

@inproceedings{LengHetherington2013-gpuWattch,
  author = {Leng, Jingwen and Hetherington, Tayler and ElTantawy, Ahmed and Gilani, Syed and Kim, Nam Sung and Aamodt, Tor M. and Reddi, Vijay Janapa},
  title = {{GPUWattch: enabling energy optimizations in GPGPUs}},
  year = {2013},
  isbn = {9781450320795},
  publisher = {Association for Computing Machinery},
  address = {New York, NY, USA},
  url = {https://doi.org/10.1145/2485922.2485964},
  doi = {10.1145/2485922.2485964},
  booktitle = {{Proceedings of the 40th Annual International Symposium on Computer Architecture}},
  pages = {487–498},
  numpages = {12},
  location = {Tel-Aviv, Israel},
  series = {ISCA '13}
}

@inproceedings{KandiahPeverelle2021-accelWattch,
  author = {Kandiah, Vijay and Peverelle, Scott and Khairy, Mahmoud and Pan, Junrui and Manjunath, Amogh and Rogers, Timothy G. and Aamodt, Tor M. and Hardavellas, Nikos},
  title = {{AccelWattch: A Power Modeling Framework for Modern GPUs}},
  year = {2021},
  isbn = {9781450385572},
  publisher = {Association for Computing Machinery},
  address = {New York, NY, USA},
  url = {https://doi.org/10.1145/3466752.3480063},
  doi = {10.1145/3466752.3480063},
  booktitle = {{Proceedings of the 54th IEEE/ACM International Symposium on Microarchitecture}},
  pages = {738–753},
  numpages = {16},
  location = {Virtual Event, Greece},
  series = {MICRO},
  month = {10}
}

@article{faiss-gpu,
  title={{Billion-scale similarity search with {GPUs}}},
  author={Johnson, Jeff and Douze, Matthijs and J{\'e}gou, Herv{\'e}},
  journal={IEEE Transactions on Big Data},
  volume={7},
  number={3},
  pages={535--547},
  year={2019},
  publisher={IEEE}
}

@misc{qwen1.5-hf,
  title        = {{Qwen1.5-MoE-A2.7B: A Mixture of Experts Transformer Model}},
  author       = {{Qwen}},
  year         = {2024},
  howpublished = {\url{https://huggingface.co/Qwen/Qwen1.5-MoE-A2.7B}},
  note         = {Accessed: 2026-01-10. License: tongyi-qianwen. A transformer-based MoE decoder-only language model with 14.3B total parameters and 2.7B activated parameters.},
}

@article{Arlot_2010,
   title={A survey of cross-validation procedures for model selection},
   volume={4},
   ISSN={1935-7516},
   url={http://dx.doi.org/10.1214/09-SS054},
   DOI={10.1214/09-ss054},
   number={none},
   journal={Statistics Surveys},
   publisher={Institute of Mathematical Statistics},
   author={Arlot, Sylvain and Celisse, Alain},
   year={2010},
   month=jan }

@online{AMD_MI355X,
  title        = {AMD Instinct™ MI355X GPUs},
  author       = {{Advanced Micro Devices, Inc.}},
  year         = {2025},
  url          = {https://www.amd.com/en/products/accelerators/instinct/mi350/mi355x.html},
  note         = {Accessed: 2026-01-13},
  organization = {AMD},
}

@online{AMD_MI210,
  title        = {AMD Instinct™ MI210 Accelerators},
  author       = {{Advanced Micro Devices, Inc.}},
  year         = {2026},
  url          = {https://www.amd.com/en/products/accelerators/instinct/mi200/mi210.html},
  note         = {Accessed: 2026-01-13},
  organization = {AMD},
}

@article{2018ReprintOM,
  title={{Reprint of: Mahalanobis, P.C. (1936) "On the Generalised Distance in Statistics"}},
  author={},
  journal={Sankhya A},
  year={2018},
  volume={80},
  pages={1 - 7},
  url={https://api.semanticscholar.org/CorpusID:239595337}
}

\received{January 2026}
\received[revised]{March 2026}
\received[accepted]{March 2026}

\end{document}